\documentclass[trans]{IEEEtran}
 \pdfoutput=1
\usepackage{amsmath,amsthm}
\usepackage{graphics} 
\usepackage{epsfig,epstopdf} 
\usepackage[export]{adjustbox}
\usepackage[dvipsnames]{xcolor}
\usepackage{amsmath,bigints,bbm} 
\usepackage{amssymb}  
\usepackage[T1]{fontenc}
\usepackage[utf8]{inputenc}
\usepackage{authblk}
\usepackage{cite}
\usepackage{cleveref}
\usepackage{caption}
\usepackage{subcaption}
\usepackage{mathtools}

\usepackage[utf8]{inputenc}
\usepackage[T1]{fontenc}

\DeclarePairedDelimiter{\ceil}{\lceil}{\rceil}
\newtheorem{theorem}{\textbf{\text{Theorem}}}

\newtheorem{proposition}{Proposition}
\newtheorem{corollary}{Corollary}

\crefrangelabelformat{equation}{(#3#1#4--#5#2#6)}
\crefname{equation}{Eq.}{Eqs.}
\Crefname{equation}{Equation}{Equations}

\newcommand*\rfrac[2]{{}^{#1}\!/_{#2}}
\newcommand{\vect}[1]{\boldsymbol{#1}}

\newtheoremstyle{problemstyle}  
        {3pt}                                               
        {3pt}                                               
        {\normalfont}                               
        {}                                                  
        {\bfseries\itshape}                 
        {\normalfont\bfseries:}         
        {.5em}                                          
        {}                                                  
\theoremstyle{problemstyle}

\newtheorem{problem}{Problem}

\begin{document}

\title{Aeronautical Data Aggregation and \\ Field Estimation in IoT Networks:\\ 
Hovering \& Traveling Time Dilemma of UAVs}

\author{  
\IEEEauthorblockN{Osama M. Bushnaq, {\em Student Member, IEEE}, Abdulkadir Celik, {\em Member, IEEE}, Hesham ElSawy, {\em Senior Member, IEEE},\\
 Mohamed-Slim Alouini, {\em  Fellow, IEEE}, and Tareq Y. Al-Naffouri, {\em Senior Member, IEEE}.} 
 \thanks{O. Bushnaq, A. Celik, M.-S. Alouini, and T. Y. Al-Naffouri  are with Computer, Electrical and Mathematical Sciences and Engineering (CEMSE) Division of King Abdullah University of Science and Technology (KAUST), Thuwal, KSA. (E-mails: {osama.bushnaq@kaust.edu.sa}, {abdulkadir.celik@kaust.edu.sa}, {slim.alouini@kaust.edu.sa}, and {tareq.alnaffouri@kaust.edu.sa})  
        
        H.\ ElSawy is with the Department of Electrical Engineering, King Fahd University of Petroleum \& Minerals, Dhahran, KSA. (E-mail: {hesham.elsawy@kfupm.edu.sa})          
	}	
}

\markboth{IEEE Transactions on Wireless Communications}{Bushnaq \MakeLowercase{\textit{et al.}}:Aeronautical Data Aggregation and Field Estimation in IoT Networks: 
Hovering \& Traveling Time Dilemma of UAVs}
\maketitle

\begin{abstract}
The next era of information revolution will rely on aggregating big data from massive numbers of devices that are widely scattered in our environment. Most of these devices are expected to be of low-complexity, low-cost, and limited power supply, which impose stringent constraints on the network operation. In this regard, this paper investigates aerial data aggregation and field estimation from a finite spatial field via an unmanned aerial vehicle (UAV). Instead of fusing, relaying, and routing the data across the wireless nodes to fixed locations access points, a UAV flies over the field and collects the required data for two prominent missions; data aggregation and field estimation. To accomplish these tasks, the field of interest is divided into several subregions over which the UAV hovers to collect samples from the underlying nodes. To this end, we formulate and solve an optimization problem to minimize total hovering and traveling time of each mission. While the former requires the collection of a prescribed average number of samples from the field, the latter ensures for a given field spatial correlation model that the average mean-squared estimation error of the field value is no more than a predetermined threshold at any point. These goals are fulfilled by optimizing the number of subregions, the area of each subregion, the hovering locations, the hovering time at each location, and the trajectory traversed between hovering locations. The proposed formulation is shown to be {\em NP-hard mixed integer} problem, and hence, a decoupled heuristic solution is proposed. The results show that there exists an optimal number of subregions that balance the tradeoff between hovering and traveling times such that the total time for collecting the required samples is minimized. 
\end{abstract}

\begin{IEEEkeywords}
Unmanned aerial vehicle (UAV); internet of things (IoT); stochastic geometry; coverage problem; aerial field estimation.
\end{IEEEkeywords}

\section{Introduction} \label{Intro}

The Internet of Things (IoT) is a foundational building block for the upcoming information revolution and imminent smart-world era. Particularly, the IoT bridges the cyber domain to everything and anything within our physical world (e.g., goods, appliances, vehicles, light poles, parking meters, plants, etc.), which enables unprecedented ubiquitous monitoring and smart control. For a cost-effective materialization of such vision,  the IoT relies on low-cost wireless sensor nodes with short transmission ranges, limited energy supply, and constrained computational capabilities~\cite{Iot2015}. The high density and wide-spatial distribution of the IoT sensors along with the stringent operational constraints for each sensor render the conventional data aggregation and field estimation schemes (i.e., clustering and multi-hopping) obsolete~\cite{Kamal}. 

Exploiting the significant advances in global unmanned aerial vehicle (UAV) market, flying data aggregators can be used to collect the IoT big data~\cite{Zeng2016May}. Instead of disseminating and routing the data across the network to fixed location access points, UAVs with wireless communication capabilities fly across the network to collect the data. This offloads the data aggregation burden from the IoT sensors to the UAVs, which offers multifold gains for the IoT implementation and operation. For instance, recharging and maintaining the UAVs that eventually fly back to central headquarters impose much less overhead than recharging and maintaining, respectively, the sensors and access points in the field. The sensors become irresponsible for data relaying and routing, and hence, simpler and lower cost sensors can be utilized. On-demand wake-up schemes upon data collection can be utilized, which reduces energy consumption and prolongs the network lifetime.  

The recent advancement, along with the reduced cost, of UAVs have attracted much interest from the information and communication technology (ICT) industry. For instance, flying base stations can be used to improve cellular coverage, provide wireless communications during natural disasters and public safety operations, and extend wireless services for rural areas~\cite{LTE_Sky, Public_UAV}. The surging UAV use cases within the ICT motivated the research community to model, assess, and develop design paradigms for hybrid terrestrial/airborne communication systems.  For instance, the studies in  \cite{khuwaja2018survey,Sun2015April,Sun2015June, Akram2014Dec,Akram2014Dec2,Akram2016Dec} characterize signal propagation and fading over air-to-ground (A2G) and ground-to-air (G2A) communications links. Coverage probability and communication rates for UAV downlink networks are derived in~\cite{Dhillon}. An optimization framework for the positions and hover times for multiple UAVs serving downlink and uplink users are proposed in~\cite{Mozaffari1} and \cite{Mozaffari2017Nov}, respectively. However, the focus in \cite{Dhillon, Mozaffari1, Mozaffari2017Nov} is on the airborne to terrestrial coverage problems. Trajectory optimization to maximize data rate from specific ground nodes is considered in \cite{Esrafilian2019, Esrafilian2019SPAWC}. The coexistence of UAV communication system with current systems is discussed in \cite{Mozaffari2016D2D, Mozaffari2019}. In the context of uplink data aggregation, the throughput per pass for a UAV over a sensor network is characterized by in~\cite{MoeWin}. Trajectory and speed optimization framework for data aggregation from sensor networks are provided in~\cite{traj_UAV}. However, the studies in \cite{MoeWin, traj_UAV} are for fixed-wing UAVs (i.e., no hovering capability). To the best of the authors' knowledge, the problem of data aggregation and field estimation with rotary-wing (i.e., hovering capable) UAV from a large-scale IoT network has not been considered in the literature\footnote{Alternatively, a fixed-wing UAV can also be employed to hover by flying around a small circle centered at the designated hovering location.}.

The main contribution of this paper lies in integrating stochastic geometry, graph theory, signal processing, and optimization theory in order to investigate two prominent problems: data aggregation and field estimation in IoT networks where IoT devices are scattered according to a Poisson point process (PPP) over a finite field. Exploiting the hovering capability of the UAV, reliable transmissions of data can be attained by activating the nodes when the UAV is stationary. Particularly, the field is divided into several subregions where the UAV sequentially hovers over each subregion to collect data from the randomly scattered devices. At each hovering location, the UAV sends a universal activation/synchronization signal to all nodes covered from that hovering location. The activated nodes then send back their data via the slotted ALOHA protocol. Assuming all IoT devices have independent data to be aggregated, data aggregation problem minimizes the mission duration (i.e., total hovering and traveling time) such that a predetermined number of observations gathered from the entire network. On the other hand, field estimation problem assumes that each IoT device accurately observes a physical phenomenon  that is spatially correlated with nearby observations. Accordingly, our objective in this problem is to minimize the total mission duration by assuring that the average MSE is no more than a predefined threshold at any point in the network. To attain such goals, we optimize the following variables: i) the number of subregions, ii) the area of each subregion, iii) the hovering locations, iv) the hovering time at each location, and v) the trajectory between hovering locations. 

Accordingly, both of the problems fall within the class of mixed integer non-linear programming (MINLP) problems, which are known to be {\em NP-hard}. Contingent upon the hovering and traveling time dilemma, a decoupled suboptimal solution is proposed to handle hovering and traveling time separately. The traveling time subproblem is solved by further decoupling it into \textit{coverage problem} and \textit{traveling salesman problem.} In order to avoid calculation of traveling time for different network size and stop points, a closed form approximation is also developed based on the agility of the employed UAV. For the hovering time minimization, data transmission success probability is accurately characterized via stochastic geometry to account for the spatially random locations of the IoT devices. Based on the number of hovering points and their locations given by the traveling time problem, hovering time is minimized by means of closed-form and numerical solutions. Proposed solutions are verified by extensive numerical results which show that total time can be minimized by handling the hovering and traveling time dilemma.

Remainder of the paper is organized as follows: In Section \ref{sec:sys_mdl}, the system model is described. In Section \ref{sec:op1}, the optimization problems are stated, formulated and discussed. In Sections \ref{sec:T_{hover}}, \ref{sec:travel_time} and \ref{sec:T_{hover2}}, the hovering and traveling times are derived which then leads to a suboptimal solution. In Section \ref{sec:num_exp}, analytical and simulation results are provided. Finally, we provide our conclusions in Section \ref{sec:conclusion}.

\section{System model} \label{sec:sys_mdl}

We consider a dense IoT network that is confined within a finite region $\mathcal{A}$. The IoT devices are distributed in $\mathcal{A}$ according to a homogeneous PPP $\Psi$ with density $\lambda$, i.e., device positions are fixed at random locations.  Bearing in mind that the sensors are used for environmental sensing and field estimation, PPP is especially desirable and suitable for field estimation tasks. Let us consider, equi-dense (i.e., using the same number of sensors) wireless sensor networks whose spatial distribution follows PPP, Poisson cluster process (PCP), and grid point processes. Given that each sensor has a sensing coverage and that measurements are spatially correlated, the perfect sensor deployment model should have enough sensors to cover the entire field and provide some sensing redundancy to recover from node failures. In this respect, a clustered sensor deployment has  highly correlated measurements within the clusters and void regions where no sensors exist, which inherently yields a high field estimation error and inefficient utilization of the available sensors. On the other hand, a repulsive or a grid based sensor deployment would provide a better coverage at the expense of less fault tolerance due to the void regions. Thanks to its uniform spatial distribution of nodes, the PPP can therefore strike a proper balance between coverage and redundancy. 

A UAV, popularly known as a drone, is equipped as an airborne base station to collect data samples from the IoT nodes every time it flies over $\mathcal{A}$ to aggregate data or estimate the field. Unlike the IoT devices, UAV location is deterministic and calculated by solving a trajectory optimization problem.
Field estimation mission aims to collect necessary information from a highly dynamic IoT network for monitoring of inaccessible environments or habitats, contaminant tracking, surveillance in special zones, etc. A limited number of observations are utilized to estimate an underlying spatially correlated random field (e.g., temperature, humidity, etc.). The employed drone knows the boundaries of the field $\mathcal{A}$ but is oblivious to the exact locations and number of the IoT nodes within $\mathcal{A}$.  Hence, the objective is to minimize total mission duration ensuring that a certain number of samples, diversified as much as possible, are collected and a prescribed mean squared field estimation error is satisfied for the former and latter missions, respectively. 

\begin{figure}[t]
	\begin{center}
		\includegraphics[width=.9 \linewidth]{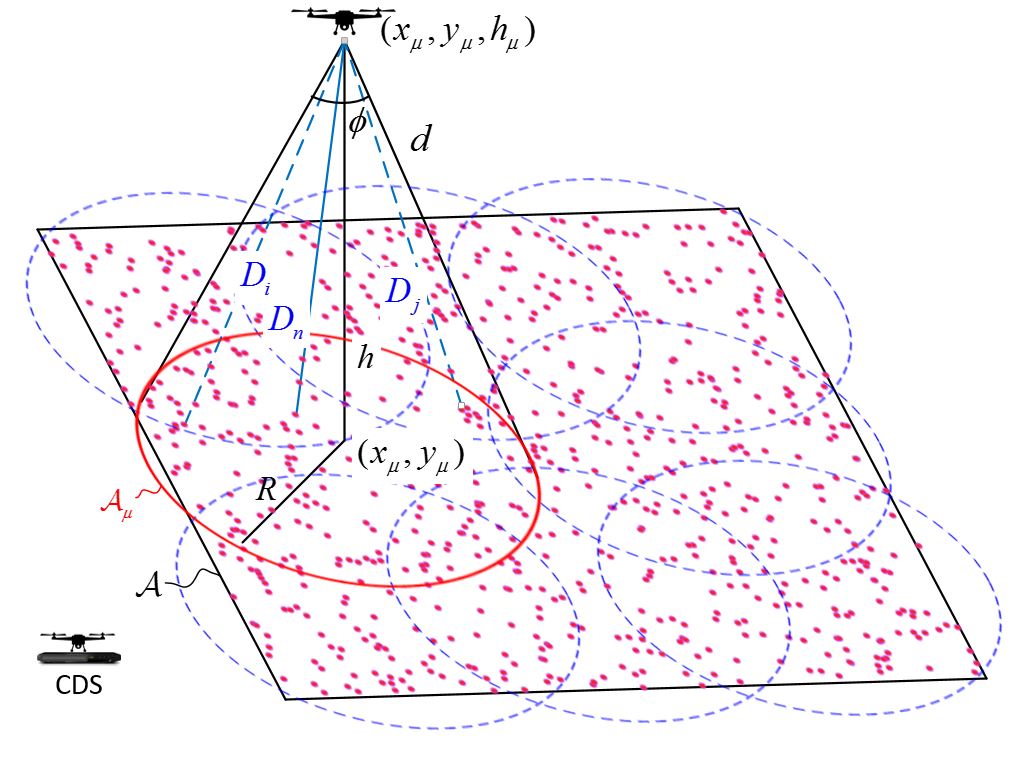}
		\caption{System model.}
		\label{fig:sys_mdl_1}
	\end{center}
\end{figure}

For each sample collection trip, the drone leaves its charging-docking station (CDS), goes to $M$ different hovering-locations (HLs) within $\mathcal{A}$, and finally returns back to the CDS. The drone stays at each HL for a deterministic time duration, denoted as the hovering time, to collect samples from the covered nodes. Fig.~\ref{fig:sys_mdl_1} shows the network model, the HLs for $M=8$, and the circular sub-regions covered from each HL.\footnote{The areas of all sub-regions, as well as the hovering time at each HL, are designed to be equal due to the spatial homogeneity of the IoT network. Otherwise, the area of each subregion (or the hovering time at each HL) should be proportional to the spatial density of the covered IoT devices.} Note that the drone's altitude $h$ for a given $M$ is determined by the radius of the circles and the antenna beam-width $\phi$ as $R=h \tan(\phi/2)$.

Given the spatial homogeneity of the IoT network, the drone stays for $T_{hover}$ at each HL such that a number of samples is collected, on average, which diversifies the acquired samples from $\mathcal{A}$. 
Let $\vect{L}= \left\{\vect{\ell_1}, \ldots, \vect{\ell_\mu}, \ldots, \vect{\ell_M} \right\}$ denote the set of HLs where $\vect{\ell_\mu}\in \mathbbm{R}^3$ is the 3D coordinates of the HL$_\mu$, $\mu \in \{1, 2, \dots, M\}$. At each HL, the covered circular region is denoted as $\mathcal{A}_\mu\left(x_\mu, y_\mu,R \right)$ with an area of $|\mathcal{A}_\mu|=\pi R^2$, where $|\cdot|$ denotes the Lebesgue measure. For a given $M$, the radius $R$ and the hovering locations $\vect{L}$ are selected such that the union of all sub-regions encompasses the entire network, i.e., $ \mathcal{A} \subseteq \bigcup_\mu \mathcal{A}_\mu$ (cf. Fig.~\ref{fig:covering_problem}). 
Furthermore, the trajectory between the HLs is selected such that the total distance traveled by the drone is minimized. It is worth noting that the drone does not move with a uniform velocity between the HLs. Instead, due to mechanical and physical constraints, the drone accelerates with rate $\hat{q}$ $[m/s^2]$ and then decelerates with rate  $\check{q}$ $[m/s^2]$ between any two stationary HLs, where the acceleration saturates at the drone maximum speed of  $\nu$ $[m/s]$. The transition from acceleration to deceleration is initiated such that the drone becomes stationary (i.e., reaches zero velocity) at the target HL.

During each hovering epoch, the drone first broadcasts a signal to activate and synchronize nodes within $\mathcal{A}_\mu$. The activated nodes transmit over a common radio spectrum using a slotted ALOHA medium access scheme with transmission probability $a$. The nodes operate at a fixed rate of $log_2(1+\beta)\, [bits/Hz/s]$, and hence, a transmission is successful if the signal-to-interference-plus-noise-ratio (SINR) at the drone is greater than $\beta$. At a given transmission instant, all activated nodes that are elected by the ALOHA protocol mutually interfere with each other. As mentioned before, the objective of data aggregation mission is to collect an average of $\zeta$ samples from $\mathcal{A}$ irrespective of the transmitting nodes' identities.  Hence, the data sent by the node with the maximum SINR is treated as the intended transmission and all other transmissions are considered as interference. 


Transmissions over the ground-to-Air (G2A) links experience power-law path-loss attenuation and Nakagami-$m$ multi-path fading. That is, the signal power decays at a rate of $D^{-\eta}$ with the distance $D$, where $\eta$ is the path loss exponent. Furthermore, the power of each transmitted signal experiences independent and identically distributed Gamma distributed gain, denoted as $G$, with the following probability density function (PDF) \cite{wackerly2002mathematical}
\begin{align}\label{eq:gammaPDF}
f_G(g) = \dfrac{m^m g^{m-1}}{\Gamma(m)} \exp(-mg),
\end{align}
where $\Gamma(\cdot)$ is the gamma function. Setting $m=1$ models  the well-known non-line-of-sight (non-LoS) Rayleigh ($m=1$) fading environment whereas $m >1$ approximates the LoS Rician fading~\cite{simon2005digital}.  For the sake of analytical tractability, the parameter $m$ is selected to be an integer, $m \in \mathbb{Z}^+$. According to the aforementioned system model, the SINR for an arbitrarily selected active node is given by
\begin{align}\label{eq:SINR}
\text{SINR}_\mu^n = \dfrac{P G_n D_{n}^{-\eta}  }{ \sum_{\bf{x} \in \tilde{\Psi}_\mu \setminus \bf{x}_n} P G_x D_x^{-\eta} + \sigma_n^2},
\end{align}
where $\tilde{\Psi}_\mu$ is the subset of nodes that are simultaneously transmitting, $P$ is the IoT nodes transmission power and $\sigma_n^2$ is the background noise variance. Note that the IoT nodes are typically manufactured with stringent power budgets, and hence, the noise variance $\sigma_n^2$ has a paramount influence on the SINR.

\section{ Hovering \&  Traveling Time Dilemma}
\label{sec:op1}

In the realm of traditional wireless networks, data aggregation is typically designed to minimize the amount and duration of data transmission in order to maximize network lifetime and utilize the available bandwidth. For a power-hungry UAV relying upon a limited battery capacity, such a goal is particularly of a vital importance in aerial data aggregation and field estimation tasks.
The hovering time of the data aggregation and field estimation missions are primarily determined by the quality of wireless links along with the number of samples to be collected and requested MSE levels, respectively. Traveling time between the HLs adds a second dimension to this already challenging task. This is because it is desirable to fulfill these tasks as soon as possible and return to the docking station for recharging.

Therefore, our main purpose in this paper is to minimize total flight duration, which can be decomposed into hovering and traveling times. Notice that minimizing the total mission duration implicitly takes the energy consumption into consideration since hovering and traveling both consume energy\footnote{Indeed, a more explicit way of minimizing the total energy consumption is considering the weighted sum of hovering and traveling time, where weights are hovering and traveling power consumption, respectively.}. At this very moment, we find ourselves on the horns of the following dilemma: On the one hand, a lower number of HLs, $M$, yields to higher latitudes and larger sub-regions, which degrades the SINR due to worse link quality and higher interference. At the extreme case of $M=1$, the drone has to hover at a sufficiently high altitude to cover $\mathcal{A}$ for a very long time to handle all traffic requests most of which have a low probability of success due to the long distance, bad channel quality, and high interference. On another hand, total traveling time increases with the number of HLs. For a high number of HLs, the drone cannot fully exploit its agility as shorter distances between the stations enforce the drone to decelerate before reaching its peak velocity. Note also that the drone needs some time for reconfiguration after arrivals to, and before departures from, HLs. More importantly, a high number of HLs may lead to unnecessarily small coverage regions that encompass no IoT devices, which wastes the traveling, hovering, and reconfiguration times related to such null HLs.
Although some studies in the literature derive a general UAV trajectory over time \cite{6,7,8} where sensor locations are known to UAVs, we intentionally modeled our problem with HLs for two reasons: 1) Sensor locations are random and unknown to the UAV, thus, the trajectory optimization formulation in \cite{6,7,8} is not applicable and 2) The data aggregation and field estimation tasks naturally results in a coverage problem. Hence, optimizing the trajectory to visit the center of each coverage regions (i.e., HLs) is inherently a traveling salesman problem. Similar to our work, Mozaffari et. al. also consider trajectory optimization across different coverage regions \cite{Mozaffari2016D2D}.  
Optimization problems of data aggregation and field estimation tasks are formulated in the following subsections.
\subsection{Data Aggregation: Collecting an Average of $\zeta$ Observations}
The formal definition of the considered aerial data aggregation problem is given as follows: 
\begin{problem}
\label{prob1}
\textit{ While ensuring that the average number of observations collected from the entire network is not less than $\zeta$, the total hovering and traveling times are minimized by optimizing the following variables: 
	\begin{enumerate}
		\item Number of HLs ($M$);\
		\item Locations of the HLs ($\bf{L}$) and radius of the circular sub-regions ($R$) such that union of the sub-regions covers all the nodes; and 
		\item Flight path of the drone which is characterized by $\vect{Z} \in \{ 0,1\}^{M \times M}$ where $z_{ij}=1$ if the drone departs from HL$_i$ and arrives at HL$_j$ and $z_{ij}=0$ otherwise.
	\end{enumerate}
}	
	\end{problem}
This problem can be formulated as in \eqref{eq:op0} where $T_{hover}^\mu$ is the hovering time at HL$_\mu$, $T_{travel}$ is the traveling time, $\vect{\varphi}$ is an auxiliary variable to exclude sub-tours in the flight paths, and $ \rm{K}_\mu$ is the number of successfully received observations at HL$_\mu$. The constraint in (3a) makes sure that the drone receives an average of $\zeta$ successful transmission from the entire network. In (3b), we guarantee that all nodes are covered by the union of the hovering areas. The constraint in (3c) ensures that each HL is visited from exactly one other HL. Likewise, (3d) assures that there is exactly one departure from each HL. In other words, ${\bf Z}$ has exactly one entry equals one in each row and column. Finally, (3e) eliminates the possible subroutes and known as Miller-Tucker-Zemlin formulation \cite{Miller1960}.
\begin{align}\label{eq:op0}
\vect{P_1}: \underset{M,R,{\bf L},{\bf Z}, \vect{\varphi}}{\min}  & \: \:T_{total} = \sum_{\mu=1}^M T_{hover}^\mu + T_{travel}, \\ 
\label{eq:op0_a} \tag{\ref{eq:op0}a)(\ref{eq:op0}b}
\hspace*{15pt} \text{s.t.}  \quad &\sum_{\mu=1}^M \mathbb{E}\{\rm{K}_\mu\} \geq \zeta , \quad \mathcal{A} \subseteq  \bigcup\limits_{\mu=1}^{M} \, \mathcal{A}_\mu, \\ 
\label{eq:op0_c} \tag{\ref{eq:op0}c} 
&   \sum_{i=1, i \neq j}^{M} {z}_{i,j} =1, \;  \forall j \in [1,M], \\
\label{eq:op0_d} \tag{\ref{eq:op0}d} 
& \sum_{j=1, j \neq i}^{M} {z}_{i,j} =1, \;  \forall {i \in [1,M]}, \\ 
\label{eq:op0_e} \tag{\ref{eq:op0}e} 
& {\varphi}_i - { \varphi}_j +M {z}_{i,j} \leq M-1, \quad 2\leq i \neq j \leq M .
\end{align} 


\subsection{Field Estimation: Ensuring an Average MSE Requirement}
 The ultimate goal of this problem is to minimize mission duration for estimating the field with an average MSE no more than a prescribed threshold, $\delta$, at any given point in the field. Assuming that the sensors perfectly measure the field values at their locations, the only source of estimation error at an arbitrary point in the field is the underlying field randomness. Although it is optimal to collect observations which are uniformly distributed as much as possible to estimate an isotropic field, successfully received observations at the UAV are generally not uniformly distributed due to the following reasons: Assuming that transmission attempts of all sensors are equiprobable (e.g., $a$), sensors located in the close proximity of the hovering circle center, $\vect{\ell}_\mu$, enjoy a higher probability of successful transmission by comparison to the sensors closer to the edge of $\mathcal{A}_\mu$. This is because sensors close to the center of $\mathcal{A}_\mu$ have a higher average GtA channel gain. This inherently implies that collecting $K$ observations from small hovering areas is more diverse and informative than collecting the same number of observations from relatively larger hovering areas. Noting that the number of required successfully received observations by the UAV varies with the number of HLs and size of the hovering circles, we define the field estimation problem in Problem \ref{prob2} and provide its mathematical formulation in \eqref{eq:op2}.
\begin{problem}
\label{prob2}
\textit{While ensuring that the average field estimation MSE at any location in the field is no more than $\delta$, the total hovering and traveling times are minimized by optimizing $M, R, {\bf L}$ and $\vect{Z}$.}
\begin{align}\label{eq:op2}
\vect{P_2}: \underset{M,R,{\bf L},{\bf Z}, \vect{\varphi}}{\min}  & \: \:T_{total} = \sum_{\mu=1}^M T_{hover}^\mu + T_{travel}, \\ \label{eq:op2_a} \tag{\ref{eq:op2}a}
\hspace*{15pt} \text{s.t.} \quad &\mathbb{E}\{ E_s\} \leq \delta , \; \forall {\bf s} \in \mathcal{A}, \quad \text{(3b)} - \text{(3e)} 
\end{align} 
\end{problem}
$\vect{P_1}$ and $\vect{P_2}$ fall within the class of mixed-integer non-linear programming (MINLP) problems which are known to be NP-Hard with a prohibitive computational complexity even for a moderate size of network. Therefore, in the remainder of the paper, we focus on developing fast yet efficient heuristic solutions. We divide the master problem into hovering and traveling time minimization sub-problems for a given M which is the most complicating (coupling) variable of $\vect{P_1}$ and $\vect{P_2}$.  Proposed sub-optimal solutions are then compared with standard benchmarks.

\section{Traveling Time}
\label{sec:travel_time}
Traveling time minimization problem can be further decoupled into two sub-problems: a \textit{coverage problem} and a \textit{trajectory optimization problem.} For a given $M$, the coverage problem seeks the minimum radius of the circles $R$ and their locations $\bf{L}$ such that the union of these circles completely covers a predefined plane. The coverage problem can be solved in two concatenated levels \cite{Nurmela2000CoveringAS,Stoyan2010}: On the inner level, $R$ is kept constant and centers of the $M$ circles are optimized to minimize the uncovered area of the sensor field. The outer level tunes $R$ whether a coverage is obtained on the inner level or not. Since the area of interest is not a constantly changing parameter, an offline table can be created for $R$ and corresponding circle locations. 

Given the HLs by coverage problem, the next step is to find the optimal flight path which minimizes the total traveling time. Indeed, the flight path can be formulated as the well-known  \textit{Traveling Salesman Problem (TSP)} which finds the shortest possible route that visits each HL and returns back to the DCS \cite{TSP1997, Kirk2014TSP}. For the optimal flight route $\vect{Z}^*$, the total flight distance is given by
\begin{align}
u= \sum_{\mu=1}^{M} u_\mu = \sum_{\mu=1}^{M} || {\bf L}_\mu - {\bf e}_\mu^T {\bf Z^*} {\bf L} ||_2,
\end{align}
where $u_\mu$ is the distance for the $\mu^{th}$ hop of the path, $|| \cdot||_2$ denotes the $\ell_2$ norm, and ${\bf e}_\mu \in \{0,1\}^{M}$ is the vector with $1$ in the $\mu$-th entry and $0$ in the rest of the entries. Using the acceleration laws, the time needed to travel distance $u_\mu$ is expressed as
\begin{align} \label{eq:tau_mu}
\tau_\mu =  \left\{ \begin{array}{cc} 
\sqrt{\dfrac{u_\mu}{\hat{q} +\check{q}}}, \quad &if \quad    u_\mu \leq \hat{u} + \check{u},  \\
\hat{t} +\check{t} + \dfrac{u_\mu - \hat{u} -\check{u}}{v},  \quad &if \quad u_\mu > \hat{u} +\check{u}, \\
\end{array} \right.
\end{align} 
where $\hat{t} = \rfrac{v}{\hat{q}} $ and $\check{t} = \rfrac{v}{\check{q}}$ are the times needed for the UAV to change its speed from $0 \nearrow v$ and from $v \searrow 0$, respectively. Similarly, $\hat{u} = \rfrac{1}{2} \hat{q} \hat{t}^2 $ and $ \check{u} = \rfrac{1}{2} \check{q} \check{t}^2$ are the required minimum distances for the drone to change its speed from $0 \nearrow v$ and from $v \searrow 0$, respectively. That is, if the distance between two hovering locations is longer than $\hat{u}+\check{u}$, the drone will be able to travel at the maximum speed after and before acceleration and deceleration steps, respectively. Finally, the overall travel time can be expressed as
\begin{align}\label{eq:t_travel}
T_{travel} = \sum_{\mu=1}^{M} \tau_\mu + M t_{conf},
\end{align}
where $t_{conf}$ is an extra time needed at each stop point for configurations after arrival and before departure, respectively.

\begin{figure}[!t]
    \begin{minipage}[b]{0.48\linewidth}
            \centering
\includegraphics[width=0.99 \linewidth]{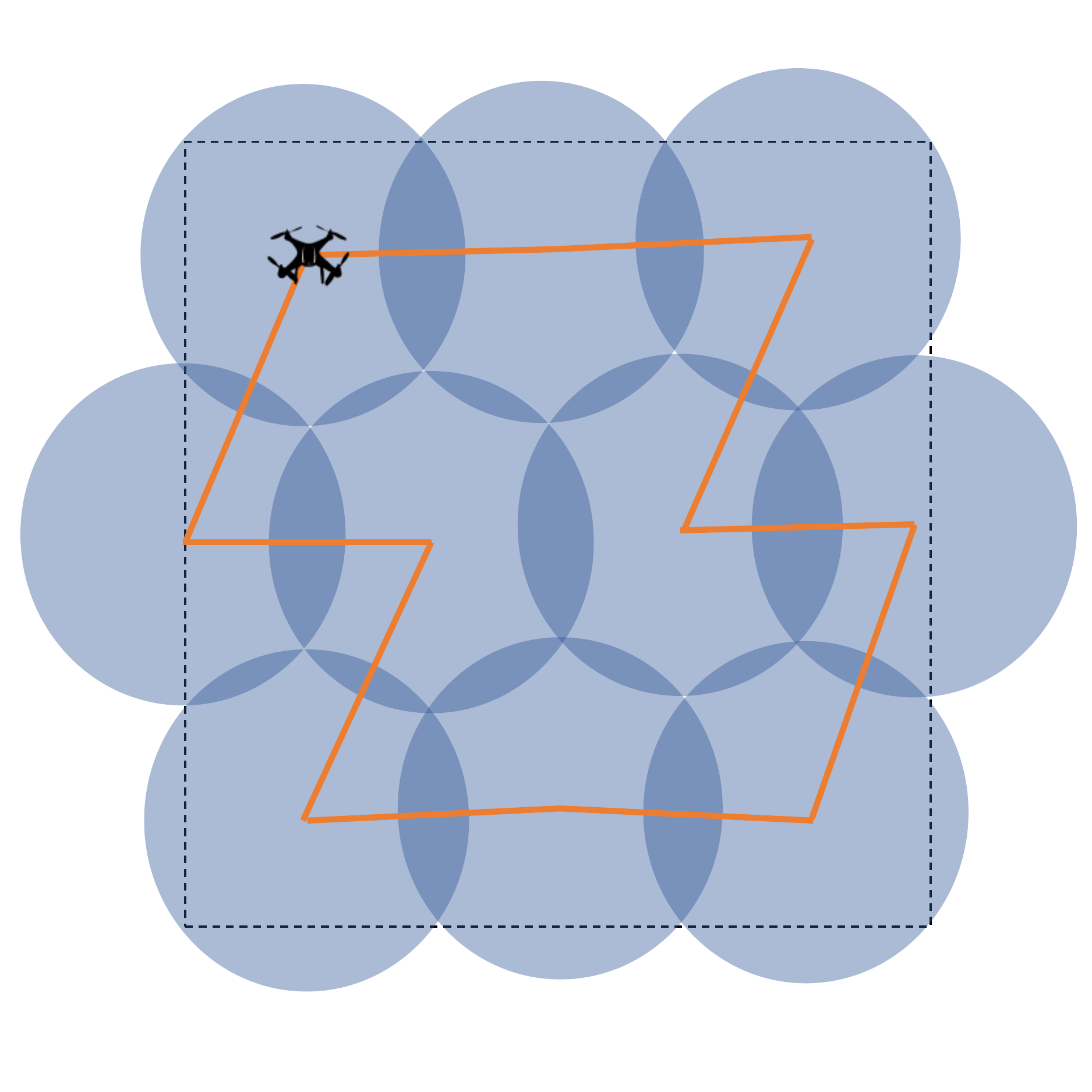}
        \caption{$M=10$}
 \label{fig:beta}
\end{minipage}
\hfill
\begin{minipage}[b]{0.48\linewidth}
            \centering
\includegraphics[width=.99 \linewidth]{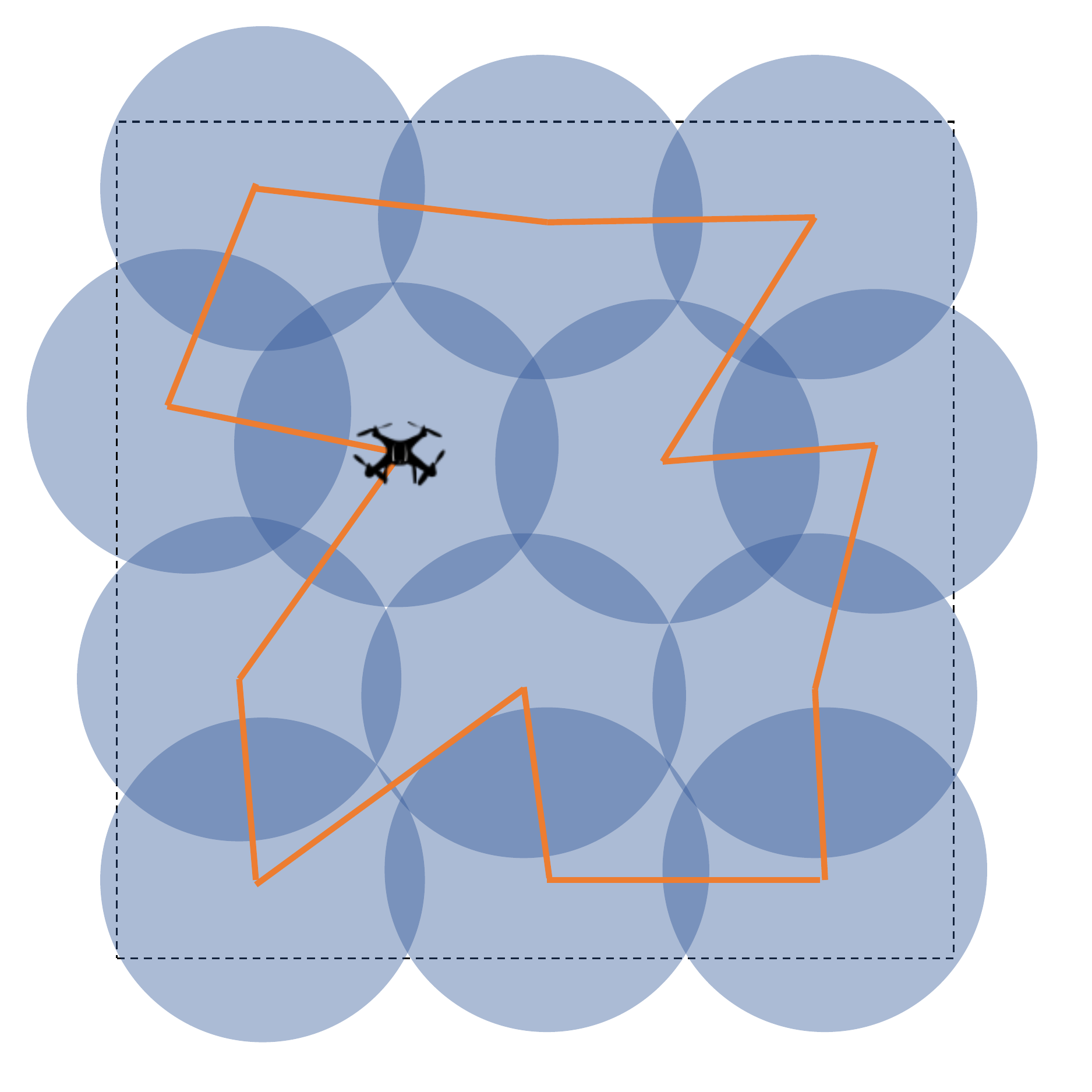}
  \caption{$M=13$}
  \label{fig:K}
\end{minipage} 
    \caption{Optimal solutions for coverage and flight trajectory problem: a) $M=7$ and b) $M=16$.  }
	\label{fig:covering_problem}
\end{figure}

%

It is important to note that the coverage problem is the main sub-problem which couples the hovering and traveling time. Therefore, for a given $(M, R, \vect{L})$, the hovering time and traveling time are two independent (decoupled) problems which do not influence each other. For a constant $\mathcal{A}$, it is quite practical to create an offline table that contains all HLs, distances, and travel times for a certain range of $M$ \cite{Bushnaq2018Aerial}. This technique simplifies the search for the optimal $M$ resulting in a minimum total hovering and traveling time. For example, Fig. \ref{fig:covering_problem} demonstrates the optimal solutions for coverage and TSP problems with $M = 9$ and $M=13$. In order to avoid computational complexity, it is necessary to develop a generic closed-form approximation of traveling time which is applicable for different scales of the network size and number of HLs. We note that the HL coordinates and radius of the coverage area are scalable with the area of the field. Therefore, we denote $\alpha_M \triangleq \frac{u}{\sqrt{|\mathcal{A}|}}$ and $\Delta_M \triangleq \frac{R}{\sqrt{|\mathcal{A}|}}$ as the total traveling distance and hovering area radius normalized by the side length of  a squared network area $|\mathcal{A}|$ for a given $M$, respectively. For a large network area such that the distances between HLs are long enough for a drone to reach its maximum speed, $v$, total traveling time is obtained by using the second case of \eqref{eq:tau_mu} as, 
\begin{align}
\label{eq:trav_approx}
T_{travel} =  \dfrac{\alpha_M \sqrt{|\mathcal{A}|}-M
(\hat{u} +\check{u})}{v}+M \tilde{t},
\end{align}
where the first term corresponds to time spent for traveling at the maximum speed while the second term accounts for total acceleration, deceleration, and configuration times between consecutive HLs, i.e, $\tilde{t} = \hat{t} +\check{t}+ t_{conf}$. Accordingly, minimum hop distance should fall within the following range,
\begin{align}
\label{eq:u_lower_bound}
\dfrac{\alpha_M \sqrt{|A|}}{M}\geq \min_{M\geq \mu \geq 1} \left( u_{\mu} \right) \geq \dfrac{v^2}{2} \left(\dfrac{1}{\hat{q}} + \dfrac{1}{\check{q}}\right) ,
\end{align}
where the upper and lower bounds are determined based on average hop range, $u/M$, and distance required to allow the UAV to accelerate to and decelerate from its maximum speed
, respectively. Following from \eqref{eq:u_lower_bound}, the network size is considered large for given drone agility parameters if 
\begin{align}\label{eq:field_size_bound}
\sqrt{|A|} \geq \dfrac{M v^2}{2\alpha_M} \left(\dfrac{1}{\hat{q}} + \dfrac{1}{\check{q}}\right).
\end{align}
Alternatively, reverse engineering of \eqref{eq:field_size_bound} gives an idea about the required drone agility for a given network size, i.e., $\frac{2 \alpha_M \sqrt{|A|}}{M} \geq v^2 \left(\dfrac{1}{\hat{q}} + \dfrac{1}{\check{q}} \right)$. Once \eqref{eq:field_size_bound} is satisfied, the proposed approximation in \eqref{eq:trav_approx} does not depend on the network size. Therefore, the total traveling time can be obtained directly from \eqref{eq:trav_approx} for a given set of $\alpha_M$ calculations as shown in Table \ref{table:1}. For a larger $M$, $\alpha_M$ values can be obtained based on extrapolation techniques, which is investigated in Section \ref{sec:num_exp}.

\begin{table}[t]
	\caption{ Obtained $\Delta_M$ and $\alpha_M$ values for $M \in [1,16]$ and $\vert \mathcal{A} \vert=1$. } \label{table:1} 
	\begin{center}
		\begin{tabular}
			{| p{.15cm} | p{.35cm} | p{.35cm} || p{.15cm} | p{.35cm} | p{.35cm} || p{.15cm} | p{.35cm} | p{.35cm} || p{.15cm} | p{.35cm} | p{.35cm} |} 
			\hline
			$M$ & $\Delta_M$ & $\alpha_M$ & $M $& $\Delta_M$ & $\alpha_M$ & $M$ & $\Delta_M$ & $\alpha_M$ & $M$ & $\Delta_M$ & $\alpha_M$ \\ [0.5ex] 
			\hline\hline
			$1$ & $.707$ & $0$ 	   &$7$ & $.274$ & $3.26$ & $13$    & $.194$ & $3.99$ & $19$ & $.158$ &$4.44$  \\ 
			\hline
			$2$ & $.559$ & $1.00$ & $8$ & $.260$ & $2.59$ & $14$   & $.186$ &$4.05$  & $20$ & $.152$ &$4.61$ \\ 
			\hline
			$3$ & $.504$ & $1.61$ & $9$ & $.231$ & $3.17$ & $15$   & $.180$ & $4.14$ & $21$ & $.149$ & $4.86$\\ 
			\hline
			$4$ & $.354$ & $2.00$ & $10$ & $.218$ & $3.59$ & $ 16$  & $.169$ & $4.26$ & $22$ & $.144$ & $5.47$ \\ 
			\hline
			$5$ & $.326$ & $2.24$ & $11$ & $.213$ & $3.37$ & $17$   & $.166$ & $4.16$ & $23$ & $.141$ & $5.06$\\ 
			\hline
			$6$ & $.299$ & $2.36$ & $12$ & $.202$ & $3.56$  & $ 18$ & $.161$ & $4.48$ & $24$ & $.138$ & $5.26$ \\ 
			\hline                        
		\end{tabular}
	\end{center} 
\end{table}

If the battery life time of a single UAV is not sufficient to cover a large area of interest, employing multiple UAVs to cooperate on data aggregation and field estimation missions would be quite beneficial. Multiple UAVs approach is also a promising approach in case of power consuming and/or time sensitive missions \cite{Esrafilian2018}. In case of a single UAV, it is expected that the sensors at the last hovering location experience a high delay in comparison with the first hovering location. In this regard, using multiple UAVs can alleviate infeasibility caused by strict constraints on battery life and delay \cite{Wu2018, Li2017}.

\begin{figure}[t]
\centering
	\begin{minipage}{.48 \linewidth} 
		\includegraphics[width=.99\textwidth]{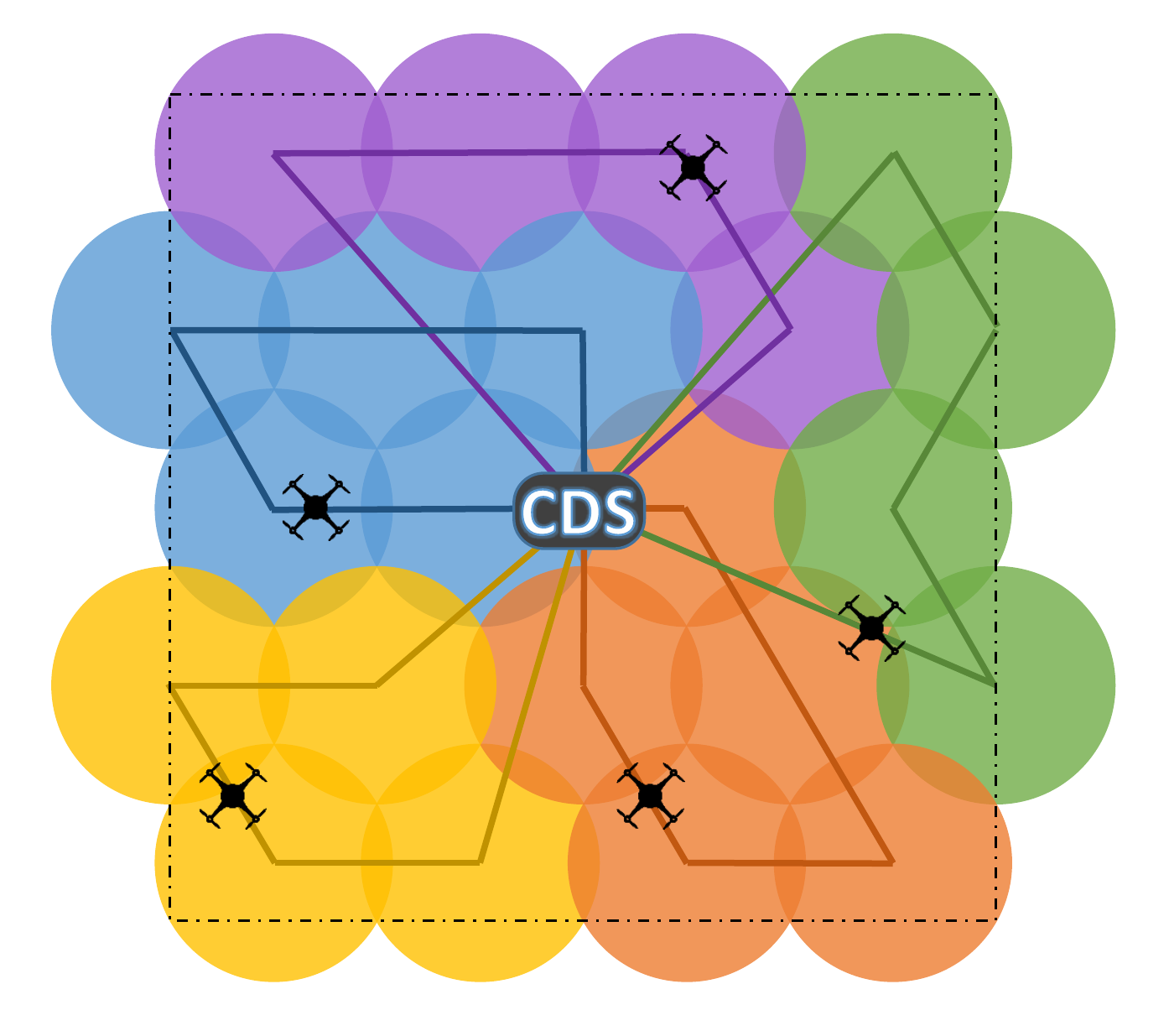}
		\caption{5 UAVs} \label{multi_UAVs_a}
	\end{minipage}
	\begin{minipage}{.48 \linewidth}
		\includegraphics[width=.99 \textwidth]{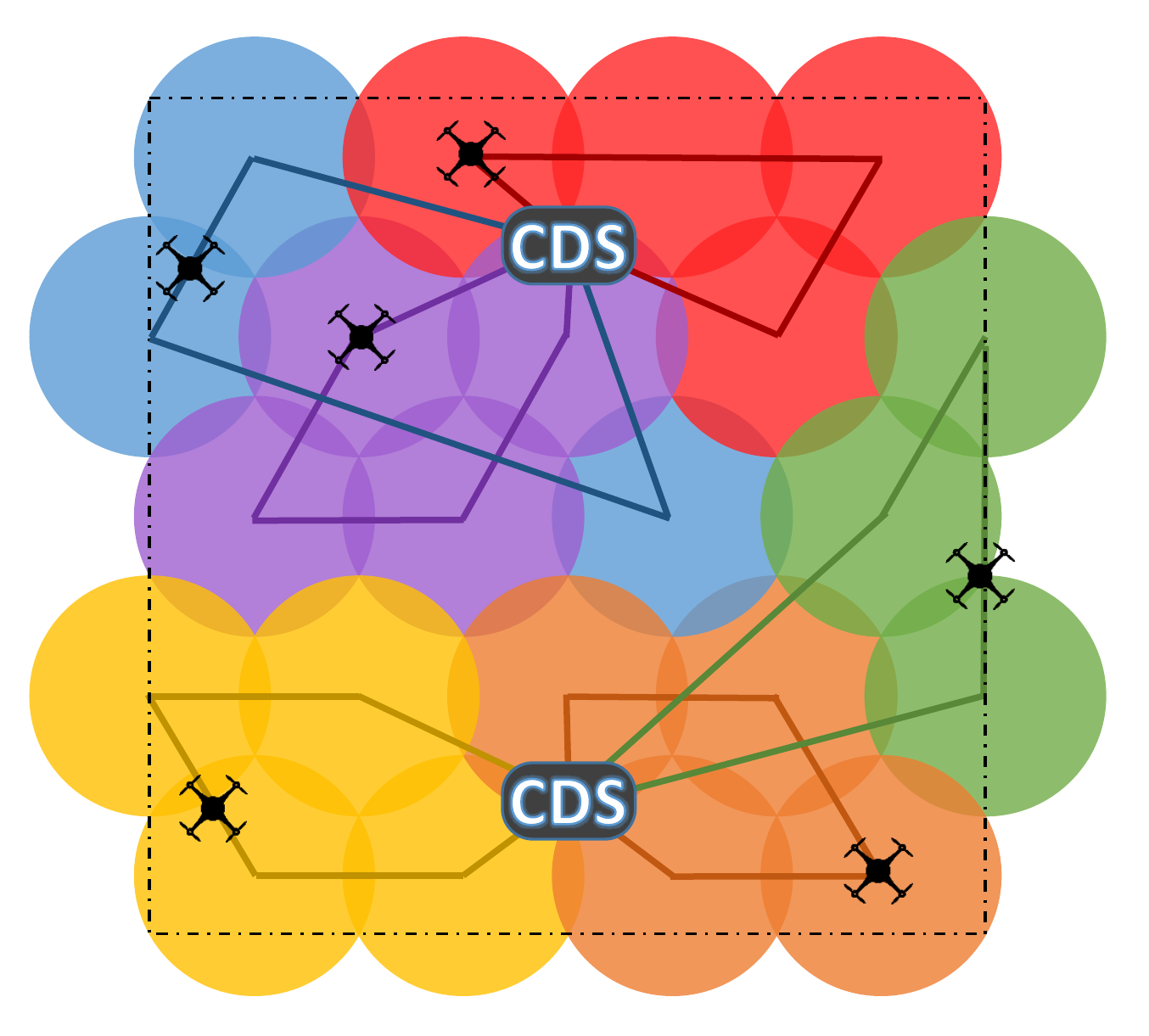}
		\caption{6 UAVs} \label{multi_UAVs_b}
	\end{minipage}
	\caption{Trajectory of Multi Cooperative UAVs.} \label{multi_UAVs}
\end{figure}

There are several techniques to approach the multiple UAVs case. One simple approach is dividing the field to several sub-fields, where each sub-field will be covered by a single UAV. Assuming identical UAVs with dedicated portions of the available bandwidth\footnote{ A similar assumption is also made in \cite{Zhang2018Analysis}. The works dealing with the multi-UAV interference can be found in \cite{Chen2018Distributed, 6}.}, which are proportional to the number of hovering locations, then the field can be divided into several equal and non-overlapping portions each covered by a single UAV. Since dedicated bandwidths eliminate the inter-UAV interference, we can apply the same methodology for each sub-field covered by a single UAV. If the UAVs have different capabilities and bandwidth allocations, the field should be divided into non-equal and non-overplaying portions, where each UAV covers a region proportional to its capabilities (e.g., speed, transmissions range, sensitivity).

Trajectory of multiple UAVs can be determined by \textit{min-max multi-depot multi-TSP ($\min\max$-MDMTSP)} approach where trajectories of $K$ UAVs can be obtained by minimizing the longest tour traveled by any UAV \cite{M_TSP, Elad2011MTSP}. By recording the  traveling time and the number of HLs per UAV separately, the total traveling time of $K$-TSP case is then expressed by
\begin{align}
T_{total} = \underset{k}{\max} \left( T_{travel}^{k} + \mu^{k} T_{hover}\right)
\end{align}
where $T_{travel}^{k}$ and $\mu^{k}$ denote the traveling time and the number of HLs for the $k$-th UAV. In order to optimize trajectories of multiple cooperative UAVs  with several CDSs, we utilize a generalized version of classical TSP, i.e.,  $\min\max$ multi-depot (e.g., CDS) multi-TSP. In Fig. \ref{multi_UAVs_a}  (Fig. \ref{multi_UAVs_b}), we show the simulation results for the sub-optimal traveling trajectories of 5 (6 UAVs) traveling over 22 HLs using single (double) CDS.

%
%
%
\section{Hovering Time for Data Aggregation}\label{sec:T_{hover}}

Hovering time is regulated by the chain relation of $M \rightarrow ( \vect{L}, R) \rightarrow\mathcal{A}_\mu $, that is, the coverage region for each HL is directly determined by $M$. We assume that the UAV utilizes an antenna with a fixed beamwidth ($\phi$) to cover a given circular geographical region \cite{Directional_Antenna, Akram2014Dec2}. For a given $M$, the height of the UAV is calculated based on corresponding hovering circle radius $R$ and $\phi$, i. e., $R = h_\mu \tan(\phi/2)$. Let ${\Psi}_\mu = \Psi \cap \mathcal{A}_\mu$ be the set of nodes covered from the HL ${\boldsymbol{\ell}}_\mu$. Then, $\#({\Psi}_\mu) \sim  {\rm Pois}(\lambda |\mathcal{A}_\mu|)$, where $\#(\cdot) $ indicates the set cardinality. Furthermore, the location of the nodes within $\mathcal{A}_\mu$ are independently and uniformly distributed. Hence, the distance between a randomly selected node in $\mathcal{A}_\mu$ and the drone is given by
\begin{align}\label{eq:distance_tx}
f_{D}(r) = \dfrac{2r}{R^2} \quad h_\mu \leq r \leq d,
\end{align}
where, as shown in Fig. \ref{fig:sys_mdl_1}, $h_\mu$ is the UAV altitude and $d= \sqrt{h_\mu^2 +R^2}$ is the hypotenuse of the triangle formed by $\vect{\ell}_\mu$, $(x_\mu, y_\mu)$, and the edge of $\mathcal{A}_\mu$. The angle between the hypotenuse and height cathetus is denoted by $\phi/2$.

The pair $( \vect{L}, R)$ has a significant impact on SINR levels and the probability of successful transmissions, which governs the hovering time to receive a target average number of successful transmissions. Assuming the ALOHA protocol for the medium access, each node independently accesses the channel and transmit to the drone with probability $a$. Let $\tilde{\Psi}_\mu \subseteq {\Psi}_\mu$ be the subset of nodes that are simultaneously transmitting to the drone at a given time instant. Then, $\tilde{\Psi}_\mu$ is a PPP with intensity function $\lambda({\bf{x}}) = a \lambda \mathbbm{1}_{\{{\bf{x}} \in \mathcal{A}_\mu\}}$, where $ \mathbbm{1}_{\{\cdot\}}$ is the indicator function that takes the value 1 if the statement $\{\cdot\}$ is true and zero otherwise. A successful sample is collected by the drone from the subset $\tilde{\Psi}_\mu$ if $\#(\tilde{\Psi}_\mu) >0$ and the received SINR at the drone is greater than $\beta$. Note that $\beta \geq 1$ implies that at most one node from $\tilde{\Psi}_\mu$ can satisfy the SINR threshold \cite{Dhillon2012}. Accordingly, we derive the probability of successful sample collection, denoted hereafter as the success probability, in the following theorem

\begin{theorem}
	\label{theorem:Ps}
	{\textit{For a given SINR threshold $\beta \geq 1$ and ALOHA transmission probability $a$, the success probability for a transmission from $\mathcal{A}_\mu$ over the G2A Nakagami-$m$ fading channel is given by
			{\small{
				\begin{align}\label{eq:P_cov}
				P_\mu^{s} &=\mathbb{P} \left(\bigcup_{\bf{x}_n \in \tilde{\Psi}_\mu} \rm{SINR}_\mu^n\geq \beta \right)= \mathbb{E}\left\{\sum_{\bf{x}_n \in \tilde{\Psi}_\mu}\mathbbm{1}_{\{\rm{SINR}_\mu^n \geq \beta\}}\right\}\nonumber\\
				&= 2 a \pi \lambda \int_{h}^{d} \sum_{k=0}^{m-1}  \dfrac{(-m\beta r^\eta)^k}{k!}  \Big[ \dfrac{\partial^k}{\partial s^k} \mathcal{L}_I(s) \Big]_{s= m\beta r^\eta} {r} dr, 
				\end{align}
			}}
			where $\rm{SINR}_\mu^n$ is the SINR when $\bf{x}_n$ is the intended transmitter and all other nodes in $\tilde{\Psi}_\mu\setminus \bf{x}_n $ are interferers, $\frac{\partial^k}{\partial s^k}$ is the $k$-th partial derivative with respect to $s$, and $\mathcal{L}_I(s)$ is the Laplace transform of the normalized interference-plus-noise power distribution, $I =  \sum_{\bf{x} \in \tilde{\Psi}_\mu \setminus \bf{x}_n} G_x R_x^{-\eta} + \sigma_n^2/P$, which is expressed as
			{\small
			\begin{align}\label{interference_laplace}
			\mathcal{L}_I(s) =
			& e^{-s \sigma_n^2/P} \exp \Bigg( -2 \pi \lambda a \int_{h}^{d} \Big( 1- \big(1+\frac{s r^{-\eta}}{m} \big) ^{-m} \Big) {r} dr    \Bigg).
			\end{align}
	}}	}
	
	\begin{IEEEproof}
		Appendix \ref{app:success}	
	\end{IEEEproof}
\end{theorem}

One can infer from Theorem \ref{theorem:Ps} that $P_\mu^{s}$ depends on the ALOHA transmission probability $a$, which should be selected carefully to maximize $P_\mu^{s}$. In conventional slotted ALOHA design, which is based on a collision model,\footnote{Collision models consider all simultaneous transmissions to be erroneous irrespective of the nodes relative channel gains. In contrast, the considered SINR capture model consider transmission success if the SINR at the receiver is greater than the threshold $\beta$ irrespective of the number of interfering nodes.}  the probability of transmission is selected such that there is only one transmission on average at every time slot. Therefore, if $N$ nodes are contending to access the spectrum, the collision based ALOHA design selects $a = 1/\rm{N}$ to minimize the collision probability. The spectral efficiency of such strategy is given by $e^{-1}\approx 36.8\%$. Note that such average single transmission per time slot may lead to a conservative channel access scheme, where several time slots may be left idle (i.e., with no transmissions).  According to the considered SINR capture model, successful transmission occurs as long as $\rm{SINR} \geq \beta$ is satisfied, which can better utilize each time slot (i.e., decrease the probability of having idle time slots) and tolerate several simultaneous transmissions.  

For the SINR capture model with a given threshold $\beta$, the success probability $P_\mu^{s}$ monotonically increases with $a$ up to a certain saddle point, then it starts to monotonically decrease as the SINR constantly degrades by allowing more transmissions. This saddle point depends on the selected  $\beta$ and can be shown to satisfy
\begin{align}
a^*= \dfrac{1}{2 \pi \lambda \int_{h}^{d} \Big( 1- \big(1+\frac{s r^{-\eta}}{m} \big) ^{-m} \Big)  r dr}.
\end{align}
which follows from the first derivative test $\frac{\partial P_{\mu}^s}{\partial a}=0$. Before proceeding to derive the hovering time, we must note that the optimal solution of $\vect{P_1}$ always requires \eqref{eq:op0_a} to hold with equality because non-strictly satisfying \eqref{eq:op0_a} (having more than $\zeta$ average successful samples) consistently requires a longer hovering time. Also considering the homogeneity of the PPP and equivalence of circular sub-regions, the drone must receive an average of $\mathbb{E}\{K_\mu\}=\zeta/M$ successful transmissions at each hovering epoch. Hence, in order to guarantee $\mathbb{E}\{K_\mu\}=\zeta/M$, the drone must hover for a duration of $J_\mu$ time slots satisfying
\begin{equation}
\label{eq:J}
J_\mu=\frac{\mathbb{E}\{K_\mu\}}{P_\mu^s}=\frac{\zeta}{ M P_\mu^s}.
\end{equation}
The time slot duration is designed in accordance with the information rate and the packet size as follows
\begin{equation}
\label{eq:tau}
\tau=\frac{S}{C}=\frac{S}{B \log_2(1+\beta)},
\end{equation}
where $S$ is the packet size, $B$ is the bandwidth, and $C$ is the Shannon capacity of the G2A channels. Collecting these expressions, we finally arrive at the following expression for the hovering time,
\begin{align}
\label{eq:T_hover}
T_{hover}^\mu = J_\mu \tau= \frac{\zeta S}{ M P_\mu^s B \log_2(1+\beta)} .
\end{align}
It is obvious from \eqref{eq:T_hover} that $\beta$ plays a critical role in the hovering duration due to the following tradeoff: The channel capacity increases with $\beta$, which yields a less transmission duration. However, an increasing $\beta$ has a negative impact on success probability, $P_\mu^s$. This motivates us to seek for an optimal SINR threshold $\beta^*=\min_{\beta}(T_{hover}^\mu)$ which could be obtained by $\frac{\partial {P_\mu^s}}{\partial \beta} =0$. Although it is not possible to derive a closed-form solution for $\beta^*$, the optimal solution can be obtained via a simple line search procedure, which is investigated in Section \ref{sec:num_exp}. 

\section{Hovering Time For Field Estimation} 
\label{sec:T_{hover2}}
In this section, we first introduce the considered spatial correlation model then obtain the hovering time for the field estimation mission. 

\subsection{Spatial Correlation Model} 
By nature, measurements of a sensor on a physical phenomenon (e.g., temperature, humidity, gas leakage, illumination, radiation, etc.) at a certain location is expected to be correlated with readings taken by other sensors within the proximity. Such correlation is captured through the covariance matrix of the field.   In this respect, the spatial correlation of the measured quantity is related to the field itself and the nature of the physical phenomena of interest. Hence, assuming any other point process does not change the spatial correlation matrix of the field. However, it may change the correlations among the reported values. At this point, it also worth mentioning that the correlations among transmissions, or better to say measured quantities, is not related to the point process.

\begin{figure}[!t]
    \centering
    \begin{minipage}{.48 \linewidth}
        \begin{subfigure}[b]{1 \textwidth}
\includegraphics[width=\columnwidth]{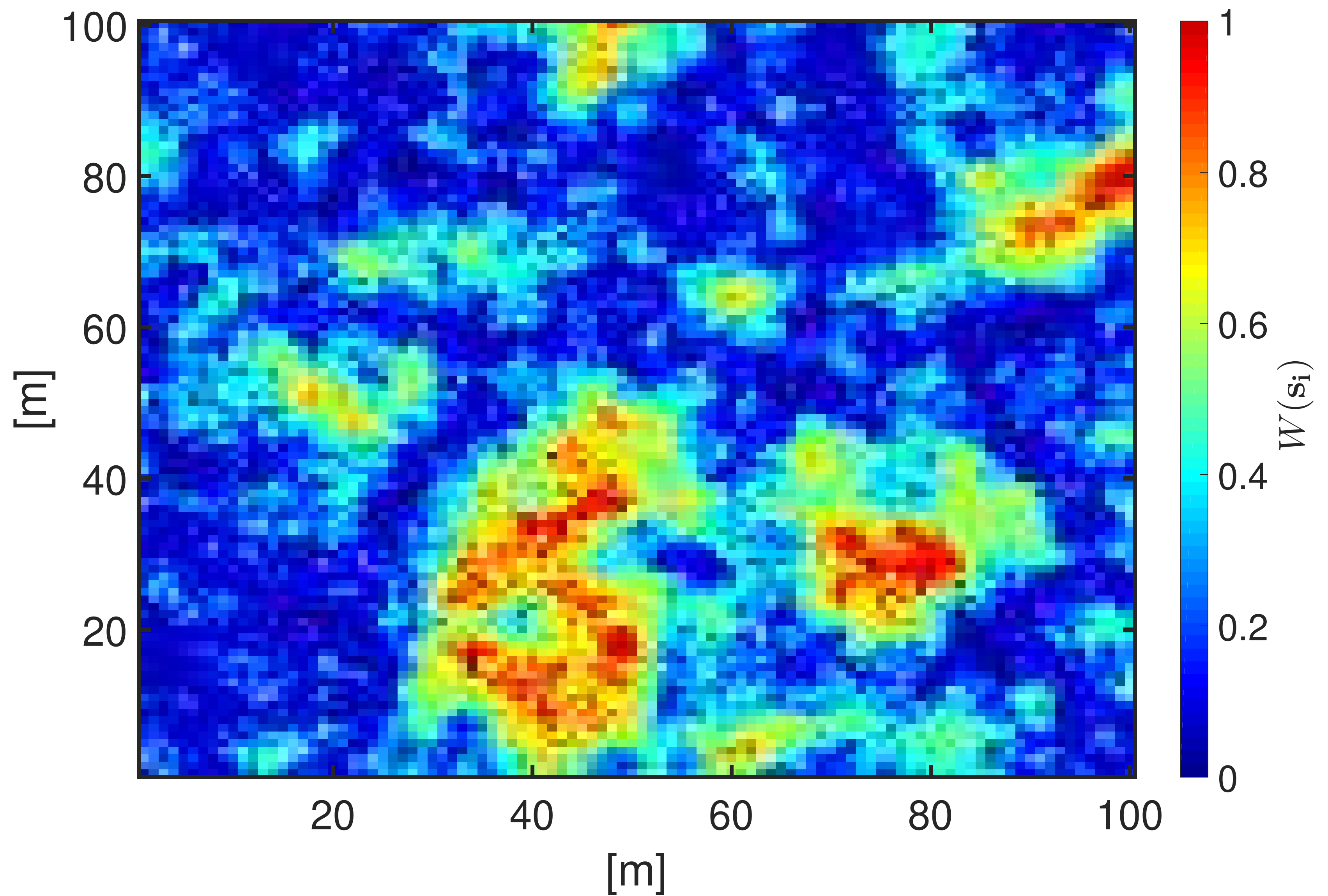}  
        \caption{Gaussian RF ($b=25$).}
    \end{subfigure}
   \begin{subfigure}[b]{1 \textwidth}
	\includegraphics[width=\columnwidth]{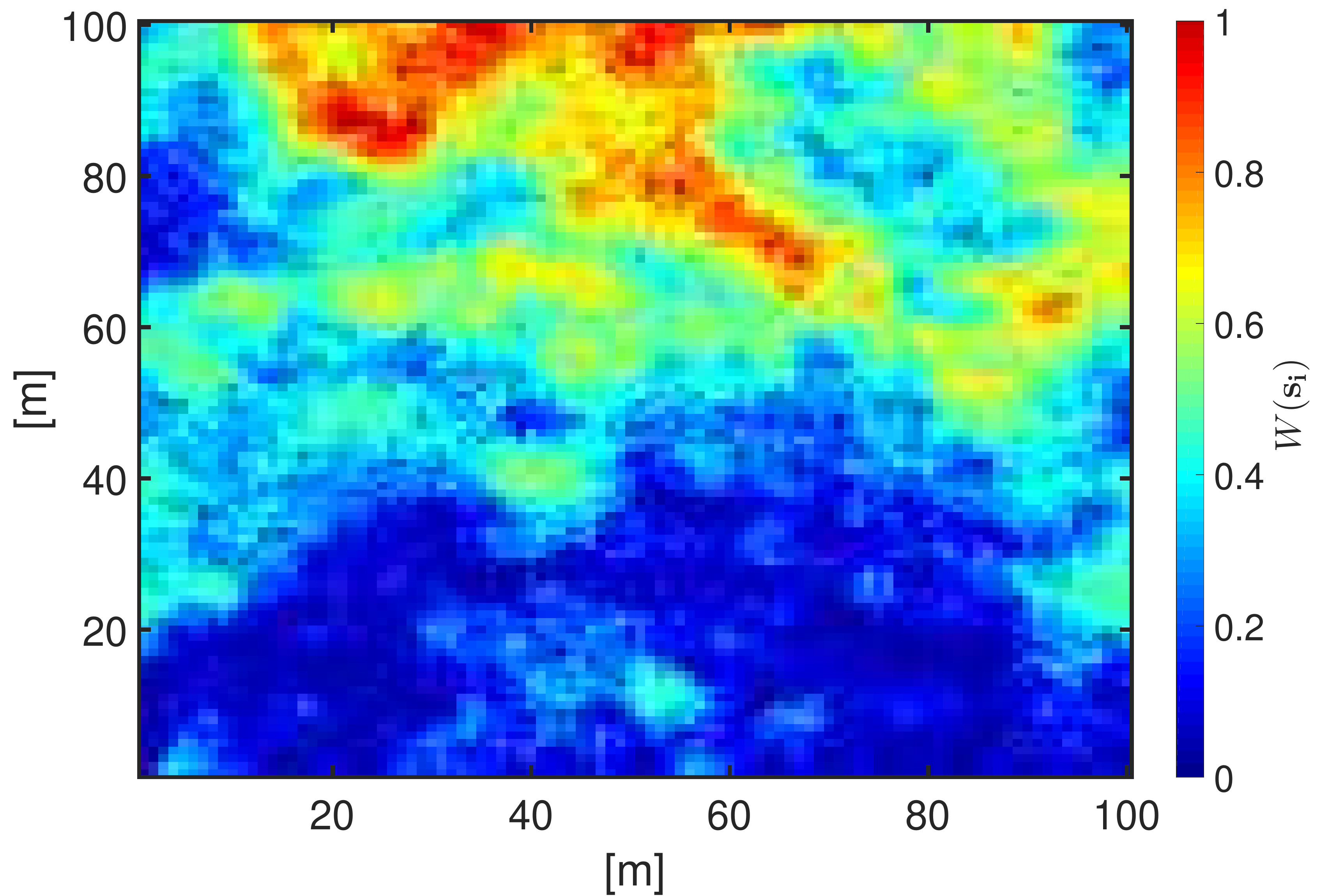}
	\caption{Gaussian RF ($b=75$).}
\end{subfigure}

    \end{minipage}
    \begin{minipage}{.48 \linewidth}
         \begin{subfigure}[b]{1 \textwidth}
 	\includegraphics[width=\columnwidth]{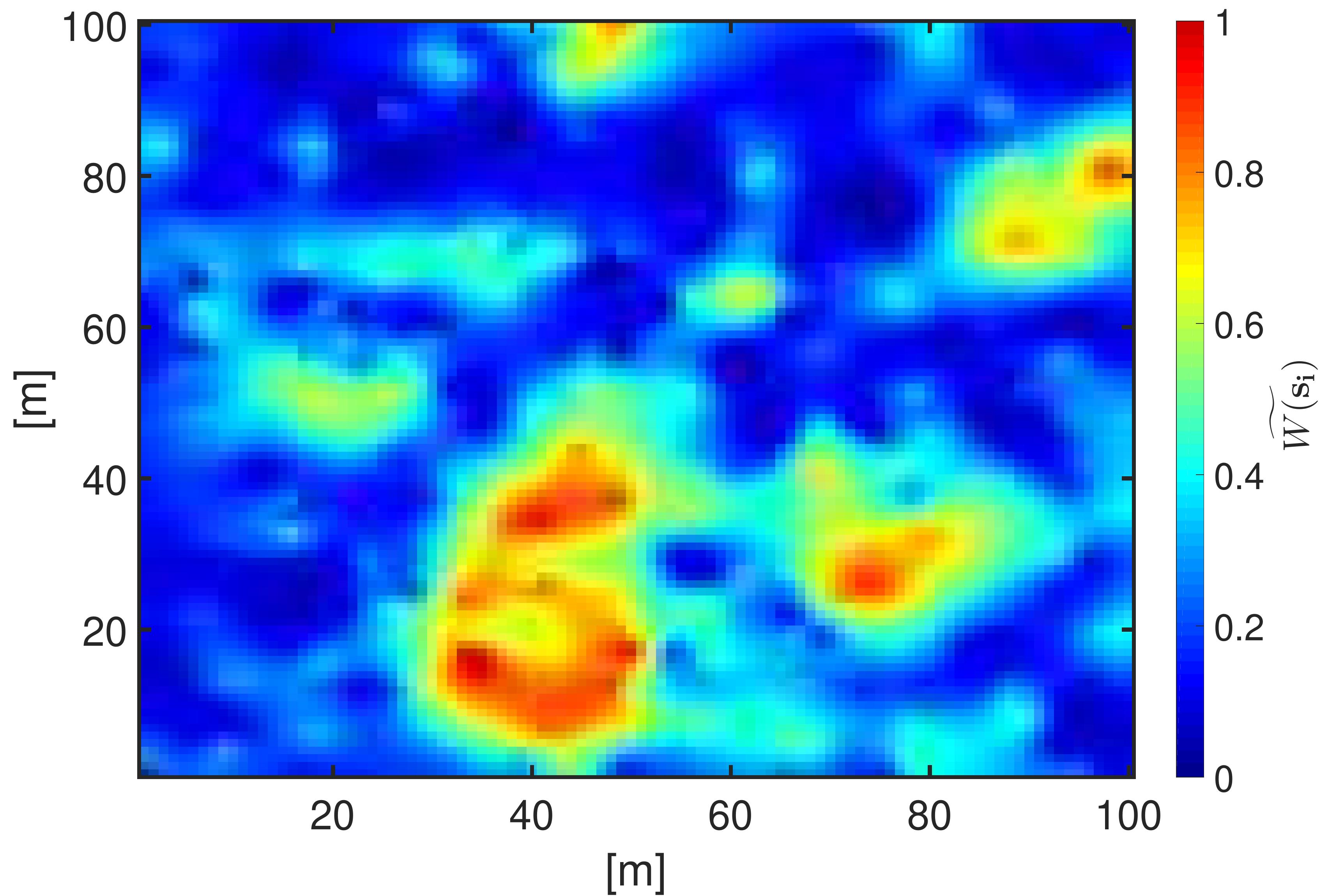}  
 	\caption{Field estimation ($b=25$).}
 \end{subfigure}   
            \begin{subfigure}[b]{1 \textwidth}
\includegraphics[width=\columnwidth]{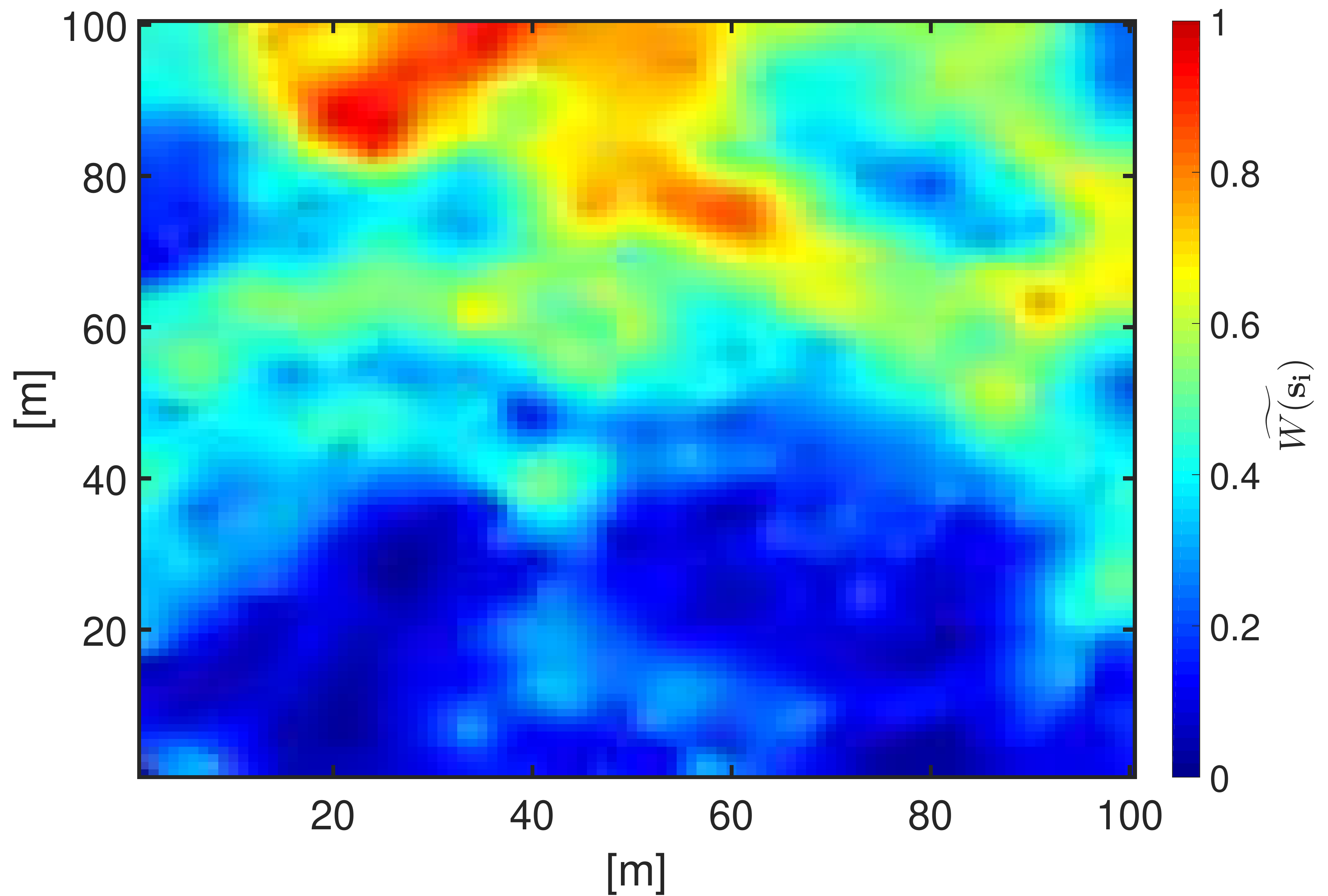}  
  \caption{Field estimation ($b=75$).}
    \end{subfigure}
    \end{minipage}
	\caption{ Gaussian random fields and corresponding estimations based on $5 \% $ random observations.}
	\label{fig:random_field_estimation}
\end{figure}

Let us consider a Gaussian random field (RF) where observations successfully received from the $i$-th sensor represent the value of the field $W({\textbf{s}_i})$ at the corresponding location ${\textbf{s}_i} \in \mathcal{A} \subset \mathbb{R}^2$. Although the Gaussian RF are mostly preferred for their analytic tractability, it may always not be suitable for all types of physical phenomena. Therefore, researchers generally mitigate its shortcomings by fitting their data to a variety of Gaussian RFs such as multi-Gaussian RF, bi-Gaussian RF, truncated Gaussian RF, skew-normal RF \cite{Marc2018,11,12,13,14}. Since the proposed model is general and is not targeting a specific application, we preferred the Gaussian model that is generic and tractable enough to understand the behavior (e.g., the interplay between hovering \& traveling time) and draw necessary design insights. Assuming a stationary isotropic covariance function, i.e., spatial correlation depends merely on the distance between sensing locations, the covariance between locations of nodes $i$ and $j$ can be expressed based on the Matérn model as 
\begin{align}
\label{eq:matern_cov}
&\Sigma({\textbf{s}}_i,{\textbf{s}}_j) = {\text{cov}}\{W({\textbf{s}}_i),W({\textbf{s}}_j)\}  \nonumber \\
&= \dfrac{\sigma^2 }{\Gamma(\nu) 2^{\nu-1}} (||{\textbf{s}}_i- {\textbf{s}}_j||_2/b)^\nu  K_\nu( ||{\textbf{s}}_i- {\textbf{s}}_j||_2/b),
\end{align}
where $||\cdot||_2$ denotes the $\ell_2$ norm, $\Gamma(\cdot)$ denotes the gamma function and $K_\nu$ is the modified Bessel function of the second kind of order $\nu$. In \eqref{eq:matern_cov}, the random field can be modeled by modifying the field variance, smoothness and range parameters $\sigma^2$, $\nu$ and $b$, respectively. Thanks to its high flexibility to cover different spatial characteristics of the random fields, Matérn covariance model has a widespread application in the literature \cite{Marc2018, Marc2014, Peter2006}.
It is worth noting that exponential covariance function is a special case of Matérn covariance model and can be  expressed as
\begin{align}
\label{eq:expo_cov1}
\Sigma({\textbf{s}}_i,{\textbf{s}}_j) = {\text{cov}}\{W({\textbf{s}}_i),W({\textbf{s}}_j)\} = \sigma^2 \exp(- ||{\textbf{s}}_i- {\textbf{s}}_j||_2/b),
\end{align}
which follows from \eqref{eq:matern_cov} by setting $\nu = 0.5$.
Examples of Gaussian random fields for unit field variance are shown in Fig. \ref{fig:random_field_estimation} along with the corresponding estimations based on randomly observing $5 \%$ of the field. 
%

Given a set of collected observations ${\bf W}_o$, the purpose of the field estimation task is to accurately estimate the random field at missing locations ${\bf W}_m$, i.e., ${\bf W}({\bf{s}}) = [{\bf W}_o \, {\bf W}_m]^T$. The estimation of the field at a missing location is expressed as,
\begin{align} 
{\widetilde{{\bf W}}}_m &= \mathbb{E}\{{\bf W}_m|{\bf W}_o\} \nonumber\\
&= \mathbb{E}\{ {\bf W}_m \} + {\boldsymbol{\Sigma}}_{m,o} ({\boldsymbol{\Sigma}}_{o,o})^{-1} ({\bf W}_o - \mathbb{E}\{ {\bf W}_o \} ),
\end{align}
where $ \mathbb{E}\{{\bf W}_m|{\bf W}_o\}$ is the conditional expectation of ${\bf W}_m$ given ${\bf W}_o$, ${\boldsymbol{\Sigma}}_{m,o}$ is the cross covariance matrix between the missing and observed field values, and ${\boldsymbol{\Sigma}}_{o,o}$ is the auto covariance matrix between the observed field values. Denoting  the prior auto covariance matrix between the missing field values as ${\boldsymbol{\Sigma}}_{m,m}$, the posterior missing values auto covariance matrix is given by \cite{schabenberger2005statistical},
\begin{align}\label{eq:error}
{\bf {E}} = {\boldsymbol{\Sigma}}_{m,m} - {\boldsymbol{\Sigma}}_{m,o} ({\boldsymbol{\Sigma}}_{o,o})^{-1} {\boldsymbol{\Sigma}}_{o,m}.
\end{align}
Noting that $ \widetilde{{\bf W}}_m$ is a function of the posterior missing values auto covariance matrix. (i.e., $\text{tr}({\bf {E}}) $ or $\det ({\bf {E}}) $), the mean squared error of the field at a given point $s$ is then given by
\begin{align}\label{eq:mse_point}
E_s = {\mathbb {E}} \{ \sigma^2 -  {\boldsymbol{\Sigma}}_{s,o} ({\boldsymbol{\Sigma}}_{o,o})^{-1} {\boldsymbol{\Sigma}}_{o,s}\},
\end{align}
where ${\boldsymbol{\Sigma}}_{o,s}$ is the correlation vector between the field values of observations and the point $s$. One can conclude from \eqref{eq:mse_point} that a more accurate estimation can be achieved by collecting more observations with higher correlation with the field value at $s$ due to a greater ${\boldsymbol{\Sigma}}_{s,o} {\boldsymbol{\Sigma}}_{o,s}$ value. Furthermore, collecting observations from spatially distant points of a stationary isotropic random field are more informative since $E_s$ is proportional to ${\boldsymbol{\Sigma}}_{o,o}$. Therefore, unlike the data aggregation task which is interested only in the number of observations, the spatial distribution of the collected observation plays a significant role in the achievable entropy at the UAV. Accordingly, \eqref{eq:mse_point} tells us that collecting observations from locations that follow a uniform spatial distribution is desirable as it minimizes the expected distance between observations (i.e., maximizes ${\boldsymbol{\Sigma}}_{o,s}$) and maximizes the distance between observations (i.e., minimizes ${\boldsymbol{\Sigma}}_{o,o}$).

\subsection{Hovering Time to Ensure an MSE of $\delta$}
\begin{figure}
	\begin{center}
		\includegraphics[width=.8 \linewidth]{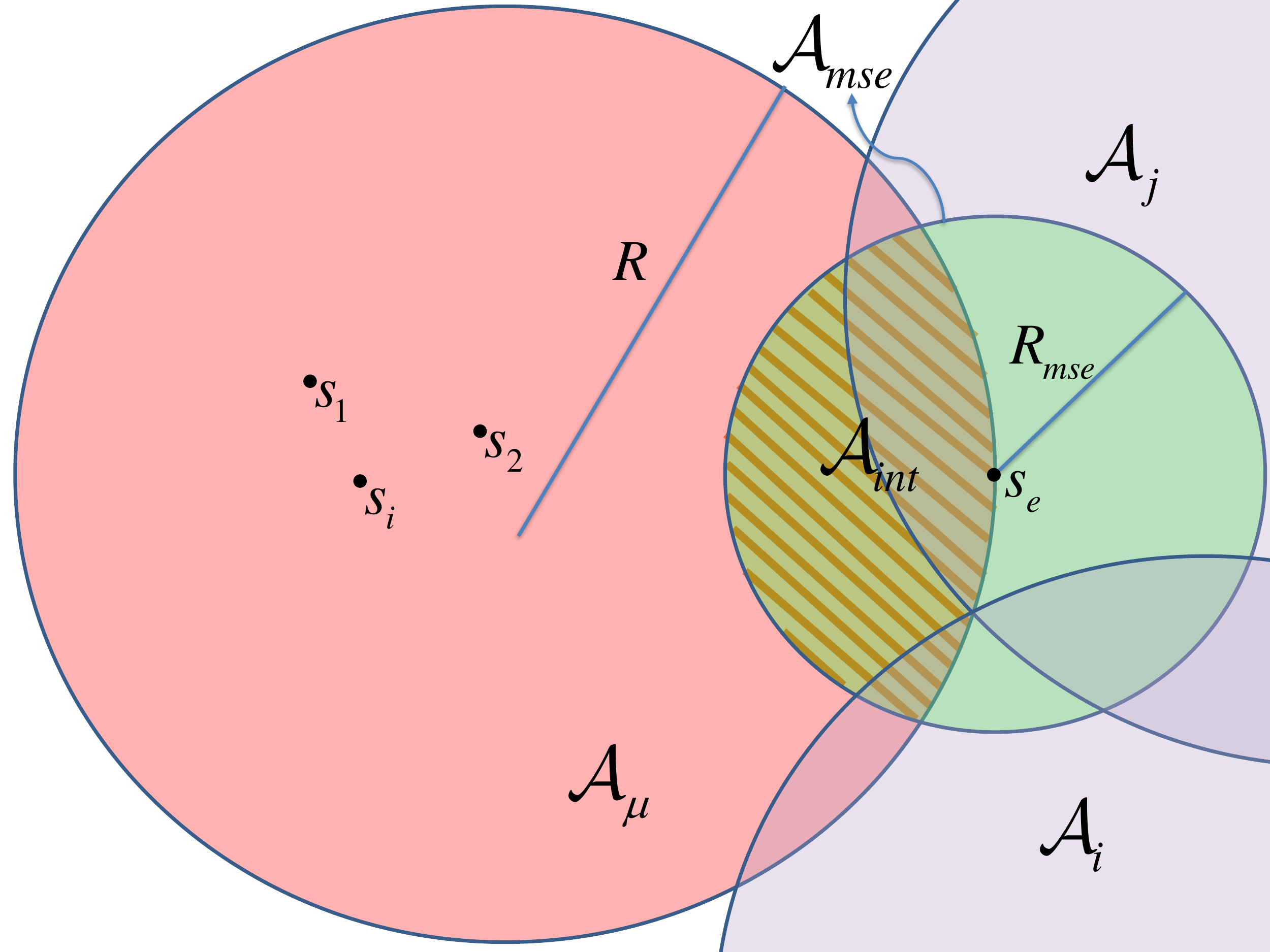}
		\caption{The illustration of hypothetical disk $\mathcal{A}_{mse}$ and its intersection with $\mathcal{A}_\mu$, $\mathcal{A}_{int}$. }
		\label{fig:hovering_area}
	\end{center}
\end{figure}

In light of above discussions, let us focus our attention on a hypothetical disk, $\mathcal{A}_{mse}$, centered at a point on the hovering disk edge with radius $R_{mse}$ as shown in Fig.\ref{fig:hovering_area}. The average SINR for a signal transmitted from a sensor closer to the center of the hovering circle is greater than that from a sensor closer to the edge. For example, a sensor located at any point inside $s \in \mathcal{A}_\mu$, has a higher probability to successfully transmit its observation to the UAV as compared to the sensor at the edge, $s_e \in \mathcal{A}_\mu$. As a result, the density of successfully received observations decreases as we get closer to the edge of $\mathcal{A}_\mu$. In other words, the average MSE is the highest at the edge of the hovering area, i.e., $\mathbb{E}\{E_s\} \leq \mathbb{E}\{E_e\}, \forall s \in \mathcal{A}_\mu$. 

In order to guarantee $\mathbb{E}\{E_s\} \leq \delta$, $\forall s\in\mathcal{A}$,  the rest of this section therefore concentrates on the worst case MSE scenario which happens at the edge of $\mathcal{A}_\mu$. To simplify the analysis of the average estimation MSE at the edge of the hovering area, we bound $\mathbb{E}(E_e)$ based on the probability of having no successful transmission from a sensor within $\mathcal{A}_{mse}$ over $J_\mu$ transmission attempts at each HL as in the following proposition. 

\begin{proposition}
\textit{For a given $\mathcal{A}_{mse}$, the average estimation MSE of a Gaussian field with the exponential covariance function is bounded by
	\begin{align}\label{eq:mse_inequality}
	\mathbb{E}\{E_e\} \leq & \sigma^2 P_{e}^{n-s}(R_{mse},J_\mu)  + \nonumber \\
	&(1-P_{e}^{n-s}(R_{mse},J_\mu)) \left(\sigma^2 - \dfrac{\exp(-2R_{mse}/b)}{\sigma^2}\right),
	\end{align} }
where $ P_{e}^{n-s}(R_{mse},J_\mu)$ is the probability of no successful transmission from a sensor within $\mathcal{A}_{mse}$ over $J_\mu$ transmission attempts. 
	\begin{IEEEproof}	
	 Appendix \ref{proposition}.	
	\end{IEEEproof}
\end{proposition}
In \eqref{eq:mse_inequality}, the first term of RHS accounts for the estimation MSE in the case of no successful observation is received from $\mathcal{A}_{mse}$, while the second term corresponds to having at least one successful transmission. Since some portions of $\mathcal{A}_{int}$ can also be covered by other hovering circles as shown in Fig. \ref{fig:hovering_area}, it is possible to receive some successful transmission from subsets intersect with hovering circles other than $\mathcal{A}_\mu$. Nonetheless, we content ourselves with the worst case scenario assuming that $\mathcal{A}_{int}$ is covered only by $\mathcal{A}_\mu$. Notice that $P_{e}^{n-s}(R_{mse},J_\mu) \to 0$ as $J_\mu \to \infty$ for any $R_{mse} > 0$. Even for $J_\mu \to \infty$, there is no guarantee to satisfy $\mathbb{E}\{E_e\} \leq \delta$ if $R_{mse}$ is set to a very large value. Hence, the size of $\mathcal{A}_{mse}$ must be carefully chosen for a minimal $J_\mu$ subject to  the MSE constraint, $\mathbb{E}\{E_e\} \leq \delta$. By substituting $P_{e}^{n-s}=0$ into \eqref{eq:mse_inequality} and rewriting for $R_{mse}$, we obtain the following upper bound
\begin{align}
R_{mse} &\leq \dfrac{b}{2}\ln\left(\dfrac{1}{(\sigma^2-\delta)\sigma^2}\right).
\end{align} 
Accordingly, the size of $\mathcal{A}_{mse}$ can be chosen by solving the following optimization problem
\begin{align}\label{eq:op3}
J_\mu^* = \quad &\underset{R_{mse}}{\min}   \quad J_\mu 
\quad s.t. \quad  \mathbb{E}\{E_e\} \leq \delta,\\
 \quad  &R_{mse} \in \left( 0, \dfrac{b}{2} \ln \left( \dfrac{1}{(\sigma^2-\delta)\sigma^2} \right ) \right).
\label{eq:op3_a} \tag{\ref{eq:op3}a)(\ref{eq:op3}b}  
\end{align}
In order derive an expression for $J_\mu$, we start by obtaining the probability of a successful transmission from $\mathcal{A}_{int} = \mathcal{A}_{mse} \cap \mathcal{A}_\mu$ as in the following theorem:
\begin{theorem}
		{\textit{For a given SINR threshold $\beta \geq 1$ and ALOHA transmission probability $a$, the probability of having a successful transmission from any point $s \in\mathcal{A}_{int}$ over the G2A Nakagami-$m$ fading channel is given by
				{\small
		\begin{align}\label{eq:P_cov-e}
			P_{e}^s = a\lambda \int_{h}^{d} \sum_{k=0}^{m-1}  \dfrac{(-m\beta r^a)^k}{k!}  \Big[ \dfrac{\partial^k}{\partial s^k} \mathcal{L}_I(s) \Big]_{s= m\beta r^a} r \theta(\sqrt{r^2-h^2}) dr,
		\end{align} }
		where
		\begin{align}\label{eq:theta}
			\theta(w) = \left\{ \begin{array}{cc} 
			2\pi &0\leq w \leq \max(0,R_{mse}-R) \\
			\Omega & |R_{mse}-R| \leq w \leq R_{mse}+R \\
			0 & \text{otherwise}. \\
			\end{array} \right.
		\end{align}
		with $\Omega = 2\arccos\Big(\dfrac{R^2+w^2-R_{mse}^2}{2R w}\Big)$
	}}
	\begin{IEEEproof}	
		Appendix \ref{app:p_success_e}
	\end{IEEEproof}
\end{theorem}
\begin{corollary}
	For $R_{mse} \leq R$, successful transmission probability from  $\forall s \in \mathcal{A}_{int}$ is given by,
	\begin{align}
	P_{e}^s = &2a\lambda \int_{h^2+(R-R_{mse})^2}^{d} \sum_{k=0}^{m-1}  \dfrac{(-m\beta r^a)^k}{k!}  \Big[ \dfrac{\partial^k}{\partial s^k} \mathcal{L}_I(s) \Big]_{s= m\beta r^a} \nonumber\\
	&r \arccos \left( \dfrac{R^2 +r^2-h^2 - R_{mse}^2}{2 R \sqrt{(r^2-h^2)}}\right) dr
	\end{align}
which directly follows from \eqref{eq:P_cov-e} and \eqref{eq:theta}.
\end{corollary}

Denoting the number of successfully received observations out of $J_\mu$ transmission attempts from $\mathcal{A}_{int}$ by Binomial random variable $\mathcal{I} \sim \mathcal{B}({J_\mu,P_e^s})$, the probability of having exactly $i$ successful receptions is given as
\begin{align}
P_{\mathcal{I}}(i|R_{mse},J_\mu) =  {J_\mu \choose i} ( 1-P_{e}^s )^{J_\mu-i} (P_{e}^s)^i.
\end{align}

Now let us continue with the remaining part of $\mathcal{A}_{mse}$, $\mathcal{A}_{ext}=\mathcal{A}_{mse} \textbackslash \mathcal{A}_{int}$,  which is covered by other hovering areas as shown in Fig. \ref{fig:hovering_area}. Similar to the analysis of $\mathcal{A}_{int}$, we limit ourselves to the worst case scenario where $\mathcal{A}_{ext}$ is covered by a single hovering disk other than $\mathcal{A}_\mu$. Reminding that all hovering areas are equal in size, the probability of having no successful transmission from $\mathcal{A}_{mse}$ from any hovering area can be approximated as,
\begin{align}\label{eq:P_no_success}
P_{e}^{n-s}(R_{mse},J_\mu) \approx  P_\mathcal{I}(0|R_{mse},J_\mu)^{1/\rho} = (1-P_e^s)^{J/\rho},
\end{align}
where $\rho$ is the ratio of $| \mathcal{A}_{int} |$ to $|\mathcal{A}_{mse}|$, i.e.,
\begin{align}
\rho = \dfrac{| \mathcal{A}_{int} |}{|\mathcal{A}_{mse}|} = \dfrac{\int_0^{R} r \theta(r) dr}{\pi R_{mse}^2}.
\end{align}

To this end, we must point out that $J_\mu$ is a decreasing function of the MSE and the problem defined in \eqref{eq:op3} is optimal at $\mathbb{E}\{E_e\} = \delta$. That is, strict satisfaction of the MSE constraint ($\mathbb{E}\{E_e\} > \delta$) requires more successful transmission and thus results in a longer hovering duration. By substituting \eqref{eq:P_no_success} into \eqref{eq:mse_inequality} and setting $\mathbb{E}\{E_e\} = \delta$, optimal hovering duration for a given $R_{mse}$ is given by
\begin{align} \label{eq:J_bound}
J_\mu^*(R_{mse}) = \ceil[\bigg]{ \rho \dfrac{\ln [ 1+(\delta -\sigma^2)\sigma^2 \exp(2R_{mse}/b)]}{\ln[1-P_{e}^s]} }.
\end{align} 
Notice that $J_\mu^*(R_{mse}) \to \infty$ when $R_{mse} \to 0$ and $R_{mse}  \to   \frac{b}{2}\ln \left(\frac{1}{(\sigma^2-\delta)\sigma^2} \right)$.  Therefore, $J_\mu^*(R_{mse}) $ has only one saddle point with respect to $R_{mse}$because the numerator in \eqref{eq:J_bound} is monotonically decreasing convex function and the denominator is monotonically decreasing concave function of $R_{mse}$. Since it is intractable to solve $J_\mu^*=\min_{R_{mse}}J_\mu^*(R_{mse}) $ by applying the first derivative test $\frac{\partial J_\mu^* (R_{mse}) }{\partial R_{mse}}= 0$, $J_\mu^*$ can be obtained numerically via a line search over $R_{mse} \in \left( 0, \frac{b}{2} \ln \left( \frac{1}{(\sigma^2-\delta)\sigma^2} \right) \right)$.
%
Accordingly, the hovering time per HL is directly proportional to $J_\mu^*$ as
\begin{align}
T_{hover}^\mu = J_\mu^* \dfrac{S}{B\log_2(1+\beta)},
\end{align}
which is the minimal hovering time to reach an MSE no more than $\delta$ at any point of the hovering circle. 

\section{Numerical Results}\label{sec:num_exp}
This section first verifies the developed mathematical model against independent Monte Carlo simulations. Then, design examples are presented where the total data collection time is minimized. Throughout the simulations, we consider a PPP with density of $\lambda =0.1$ node/m$^2$ that is distributed over a network area of $\mathcal{A}= 100 \times 100$ m$^2$ \footnote{$\lambda =0.1$ translates to 100,000 node/km$^2$. Indeed, from 100,000 to 1,000,000 node/km$^2$ is the range of ultra dense IoT.}. 
Unless stated explicitly otherwise, Table \ref{table:2} tabulates the default system parameters which are mainly drawn from evaluations of works in  \cite{Directional_Antenna, Atif2018EH, Marc2014, Marc2018, Peter2006, Akram2014Dec2}.

\renewcommand{\arraystretch}{0.75}
\begin{table}
	\caption{Default system parameters.}	\label{table:2} 
	\centering
	\begin{tabular}
		{| p{.4cm} | p{1.5cm}  ||  p{.4cm} | p{1.5cm} ||  p{.7cm} | p{1.5cm} |} 
		\hline
		Par. & Value & Par. & Value & Par. & Value \\ [0.5ex] 
		\hline\hline
		$P$ & $-30$ $dBm$ &$\sigma_n^2$ & $-80$ $dBm$ & $\phi$& $90^\circ$ \\
		\hline
		$\eta$ & $3$ & $B$ & $200 \: KHz$  & $S$ & $5 \: KB$ \\
		\hline
		$m$ & $\{1, 2, 3\}$ & $v$ & $ 20 \: km/h$ & $\hat{q}$ $(\check{q})$ & $10 \: km/h^2$  \\
		\hline
		$\lambda$ & $0.1$ & $\zeta$ & $ 250 $ &  $\nu$ & $0.5$ \\
		\hline
				$\sigma^2$ & $1$ & $b$ & $ 0.75 $ & $\delta$ & $0.2$  \\
		\hline
	\end{tabular}
\end{table}

\begin{figure}[t]
	\centering
	\begin{subfigure}[b]{0.5 \textwidth}
		\centering
		\includegraphics[width=.99\columnwidth]{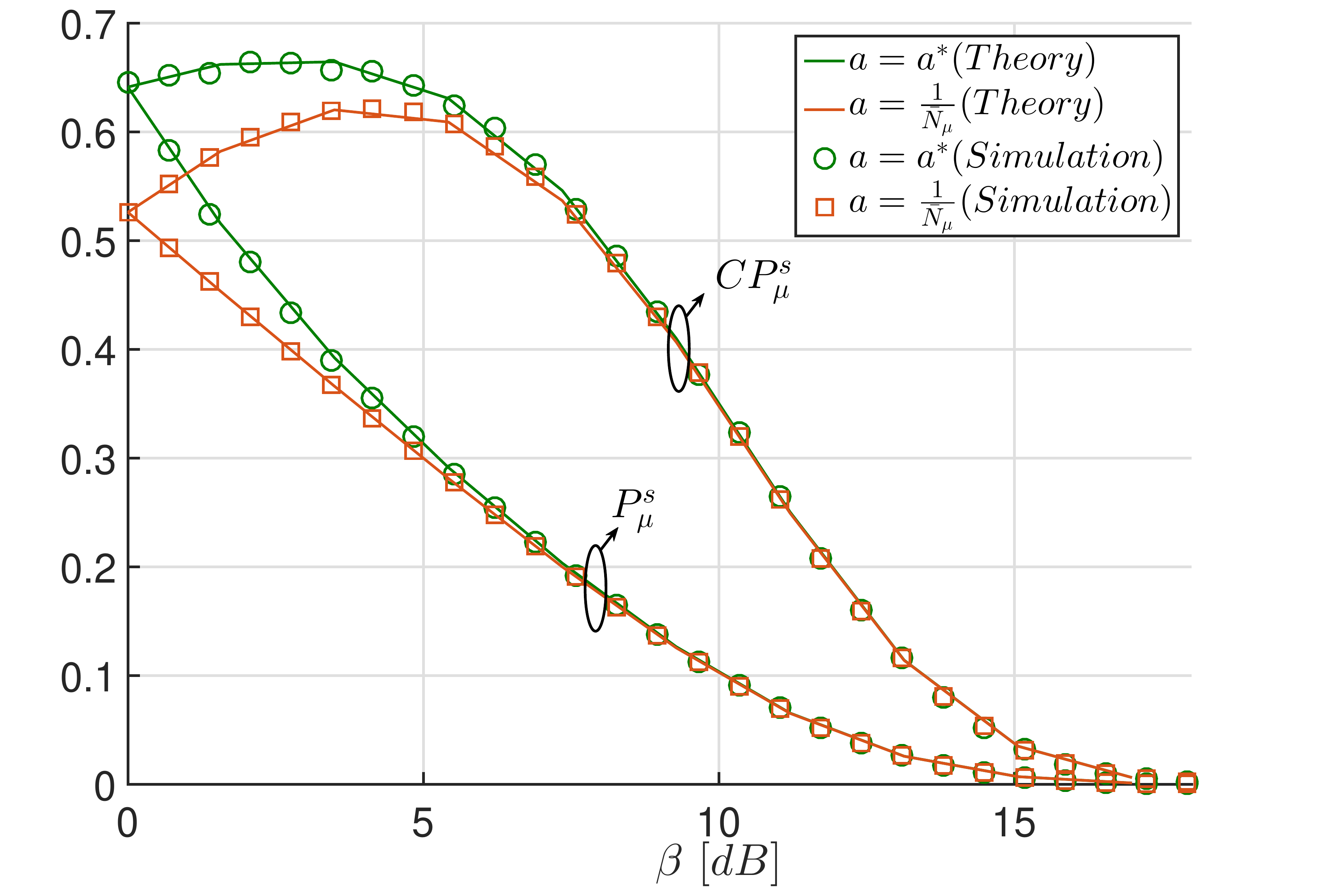}
		\caption{$P_{\mu}^s$, and {$C P_{\mu}^s$} for $a = 1/\bar{N}_\mu$ and $a= a^*$.}
		\label{fig:P_success_D_vs_beta}
	\end{subfigure}
	\begin{subfigure}[b]{0.5 \textwidth}
		\centering
		\includegraphics[width=.99 \columnwidth]{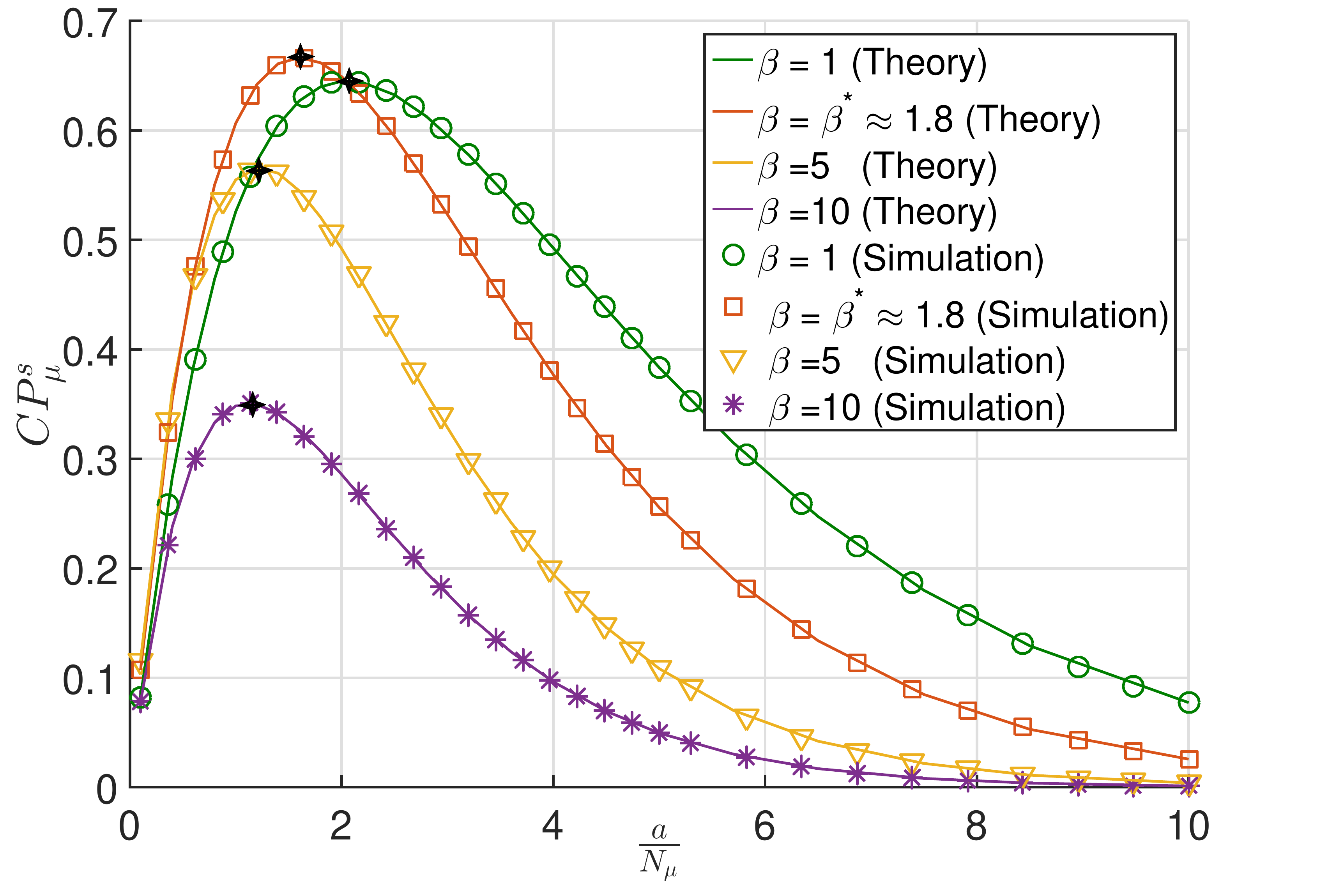}
		\caption{$CP_\mu^s$ vs. $a$ for $\beta \in\{1,\beta^* =1.8,5,10\}$.}
		\label{fig:P_success_vs_P_t}
	\end{subfigure}
	\caption{Impacts of $a$ and $\beta$ on $P_{\mu}^s$ and {$C P_{\mu}^s$}.}\label{fig:P_success}
\end{figure}

Consider a specific hovering area which is defined by $R=h=20\: m$. Fig. \ref{fig:P_success_D_vs_beta} shows the success probability $P_\mu^s$ and the throughput $CP_\mu^s$ versus $\beta$ (i.e., transmission rates) for the optimal transmission probability $a^*$ and the conventional ALOHA transmission probability $a=\frac{1}{\bar{N}_\mu}$~\cite{Roberts1975}, where $\bar{N}_\mu=\lambda \pi R^2$ is the average number of nodes in $\mathcal{A}_\mu$. The proposed optimal transmission probability $a^*$ is shown to deliver a superior performance in comparison to the conventional collision based ALOHA design. Since a higher $\beta$ implies that moderate interference levels can even lead to a failure, the performance of both schemes converges after a certain $\beta$. In such cases, an average of a single transmission per time slot is desirable, which comes at the expense of being conservative and having underutilized time slots with no transmissions.  A similar discussion also applies in Fig. \ref{fig:P_success_vs_P_t} where $C P_\mu^s$ is plotted against the ALOHA transmission probability ${a}$ and $a^*$ are shown with black colored stars. Apparently, increasing $a$ improves the success probability up to a certain extent as the time slots are better utilized with diversified transmissions. However, increasing $a$ beyond an optimal utilization point leads to an aggressive transmission scheme where interference dominates the failure probability. The impact of $\beta$ on $C P_\mu^s$ is also illustrated in Fig. \ref{fig:P_success_vs_P_t} where $C P_\mu^s$ decreases as $\beta$ moves away from the optimal point, $\beta^* \approx 1.8$. 

\begin{figure}[t]
    \centering
    \begin{subfigure}[b]{0.5 \textwidth}
    	\centering
		\includegraphics[width=.99\columnwidth]{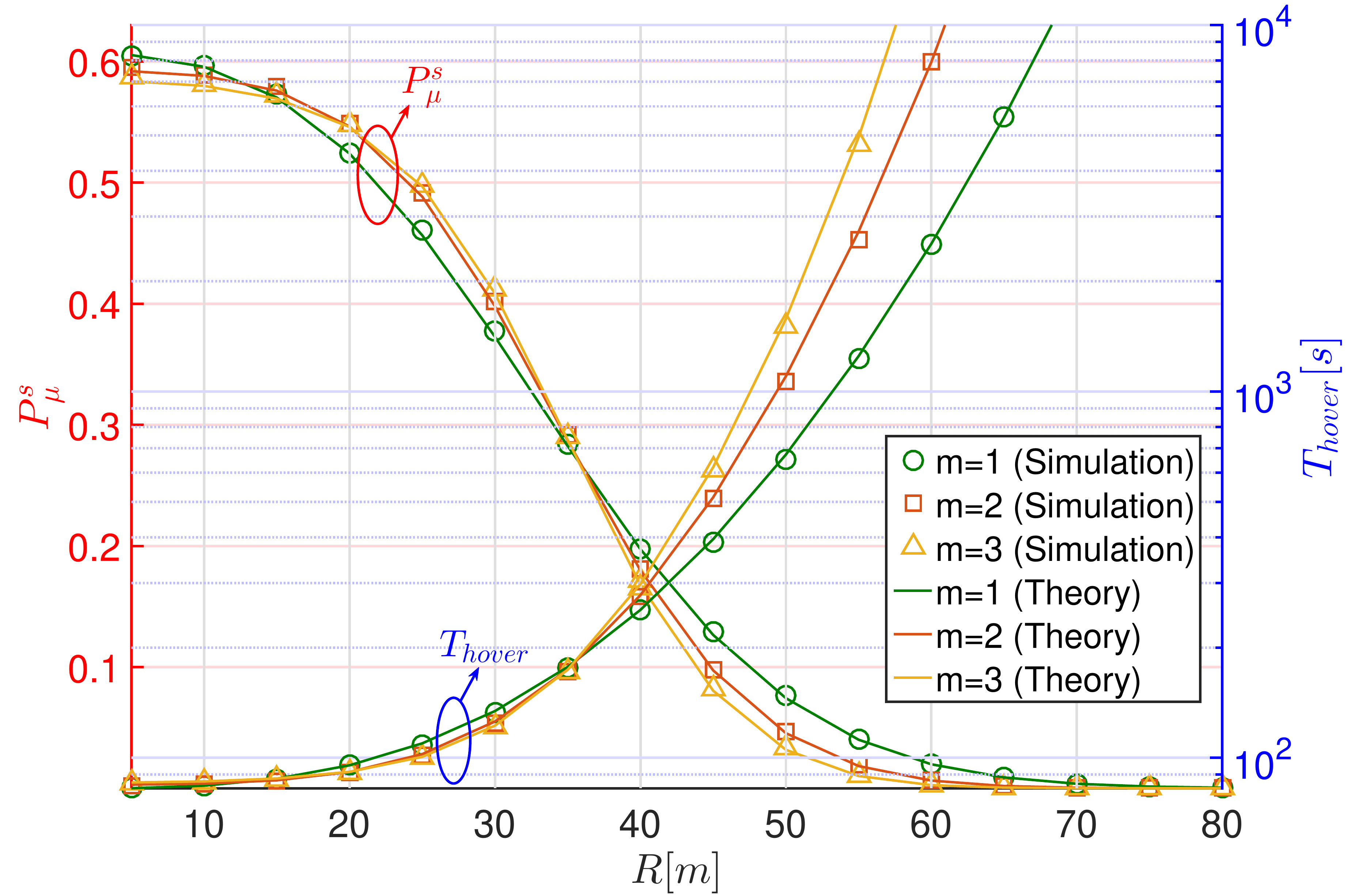}
		\caption{$P_\mu^s$ and $T_{hover}^\mu$ vs. hovering circle radius, $R$.}
		\label{fig:P_success_T_hover_vs_h}
    \end{subfigure}
    \begin{subfigure}[b]{0.5 \textwidth}
    	\centering
		\includegraphics[width=.99\columnwidth]{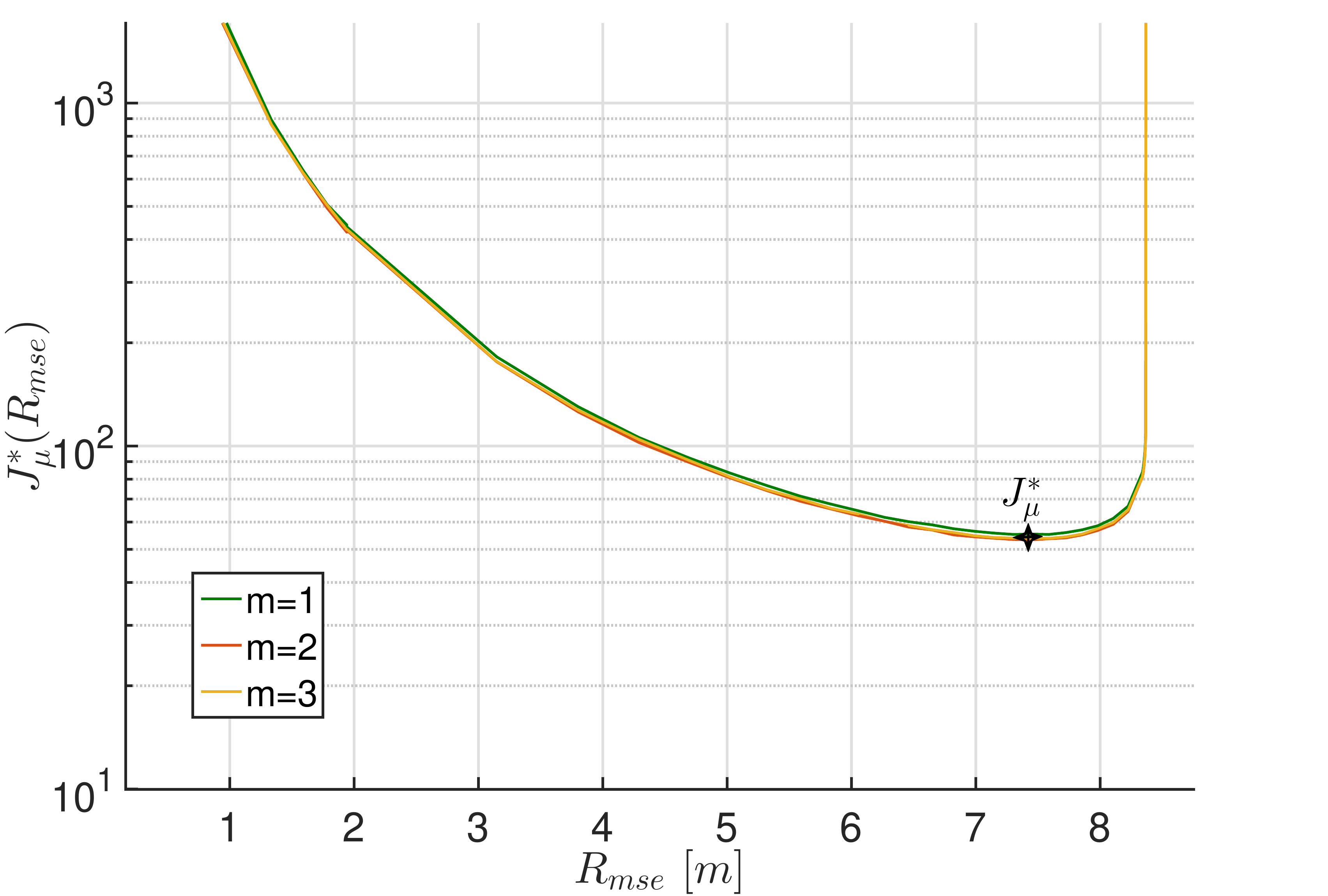}
		\caption{$J_\mu$ vs. $R_{mse}$ for $\delta=0.2$ and $\beta=1$.}
		\label{fig:J_vs_R_mse}
    \end{subfigure}
    \caption{Impacts of $R$ and $R_{mse}$ on $T_{hover}^\mu$ and $J_\mu$.}
    \label{fig:R-Rmse}
\end{figure}

The hovering and traveling time dilemma is visualized in Fig. \ref{fig:P_success_T_hover_vs_h} where $P_\mu^s$ and $T_{hover}^\mu$ are plotted versus the radius $R$, i.e., $|\mathcal{A}_\mu|$. For given $\beta$ and $a$, increasing the hovering area drastically reduces $P_\mu^s$, which directly increases the hovering time since the drone needs to wait for a longer time duration to collect an average of $\zeta/M$ successful samples. On the other hand, traveling time naturally reduces since a larger $R$ means less number of stop points, $M$. Now, let us consider a random field with $\sigma^2 = 1$ and $b=75$. Fig. \ref{fig:J_vs_R_mse} demonstrates optimal number of required transmission attempts, $J_\mu^*(R_{mse})$, to achieve an MSE no more than $\delta=0.2$ against $R_{mse}$ values. As discussed in Section \ref{sec:T_{hover2}}, $J_\mu^*(R_{mse})$ is a convex function of $R_{mse}$  , $J_\mu^*(R_{mse}) \to \infty$ as $R_{mse} \to 0$ and $R_{mse} \geq  \frac{b}{2}\ln \left(\frac{1}{(\sigma^2-\delta)\sigma^2}\right)$. The overall optimal number of transmission attempts, $J_\mu^*=\underset{R_{mse}}{\min} \; J_\mu^*(R_{mse})$, is obtained via line search and shown with a black star shape in Fig. \ref{fig:J_vs_R_mse}.

\begin{figure}[t]
    \centering
    \begin{subfigure}[b]{0.5 \textwidth}
    	\centering
		\includegraphics[width=.99\columnwidth]{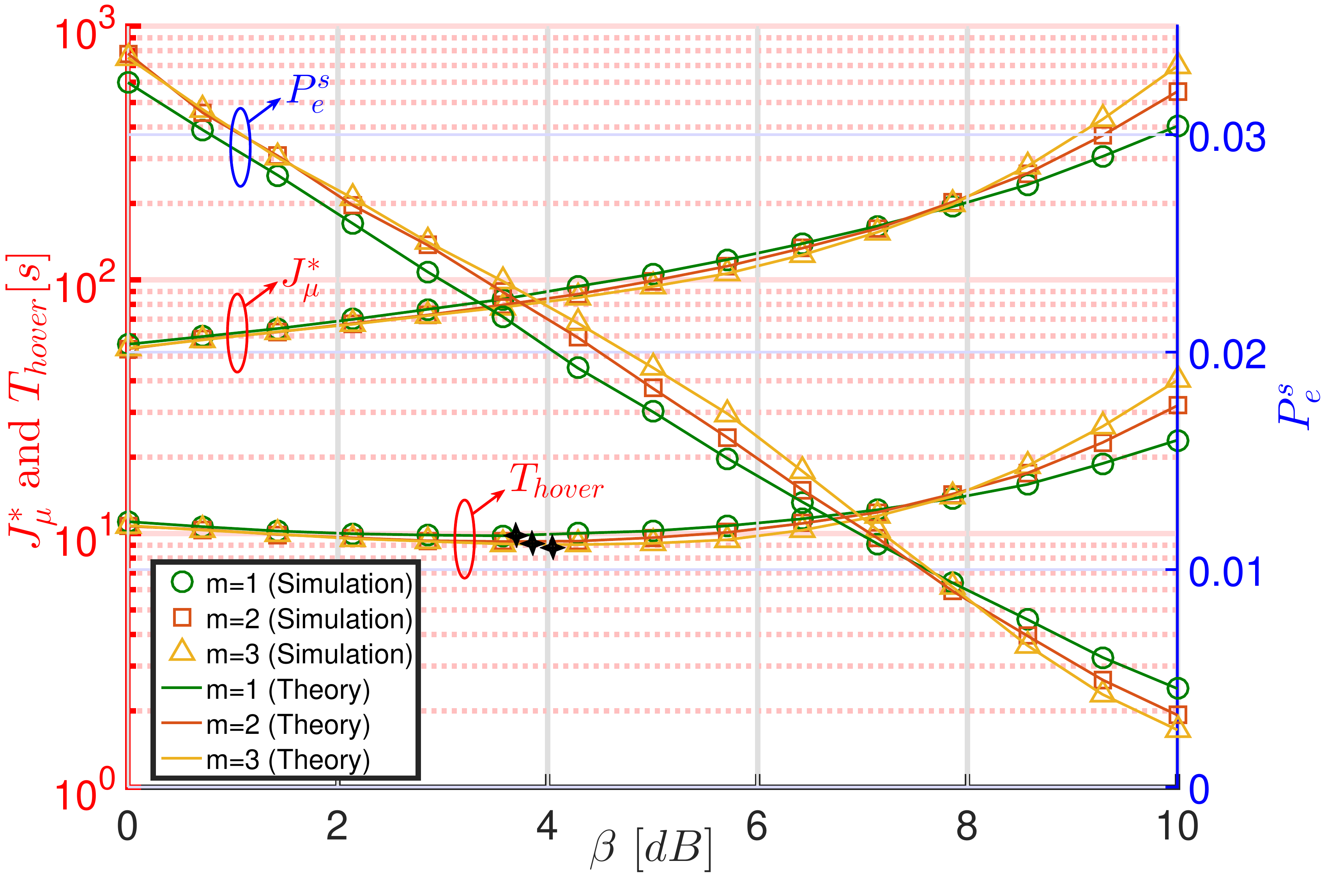}
		\caption{$\beta$ vs. $P_e^s$, $J_\mu^*$, and $T_{hover}^\mu$.}
		\label{fig:J_T_hover_vs_beta}
    \end{subfigure}
    \begin{subfigure}[b]{0.5 \textwidth}
    	\centering
		\includegraphics[width=.99\columnwidth]{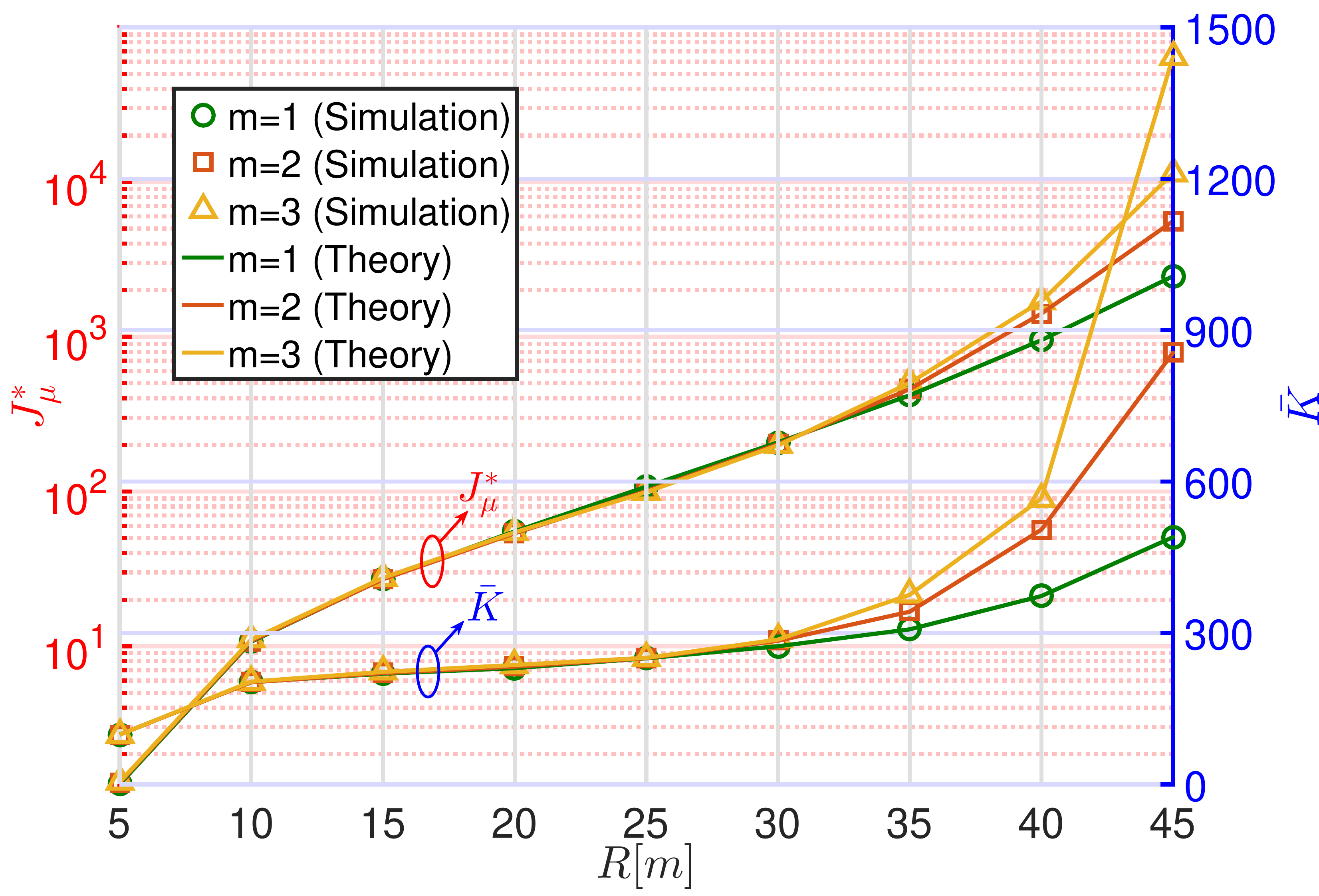}
		\caption{$R$ vs. $J_\mu^*$ and $\bar{K}$.}
		\label{fig:J_k_vs_R_c}
    \end{subfigure}
    \caption{Impacts of $\beta$ and $R$ on $T_{hover}^\mu$, $P_e^s$, $J_\mu^\star$,  and $\bar{K}$. }
    \label{fig:R-Beta}
\end{figure}

Fig. \ref{fig:J_T_hover_vs_beta} shows the behavior of success probability from $P_e^s$, the minimum number of transmissions $J_\mu^*$, and the hovering time $T_{hover}^\mu$ with respect to $\beta$. Increasing $\beta$ yields a higher channel capacity as $C=B \log_2(1+\beta$), but also decreases  $P_e^s$. Therefore, $T_{hover}$ reaches a saddle point by decreasing with increasing $\beta$, then, start increasing again. Noting that this saddle point cannot be derived analytically, we again obtain the optimal $\beta$ by a line search, which is shown by a black star mark in Fig. \ref{fig:J_T_hover_vs_beta}.

Fig. \ref{fig:J_k_vs_R_c} depicts the hovering area effects on $J_\mu^*$  and $\bar{K}$ that denotes the required total number of successful receptions to ensure an MSE of $\delta$, i.e., $\bar{K}= \mathbb{E}\{K\} = P_\mu^s \frac{J_\mu^* |\mathcal{A}|}{\pi R^2}$. At the first glance, larger hovering areas require more observations to sustain the same MSE, which naturally results in a higher $\bar{K}$. This increase is due to the fact that the distribution of successfully received observations over the field becomes less uniform as $R$ increases. One can also see that requiring a constant number of observations (i.e., $\zeta$ in the data aggregation mission) is not viable to ensure the required MSE levels. 

\begin{figure}[t]
    \centering
    \begin{subfigure}[b]{0.5 \textwidth}
    	\centering
		\includegraphics[width=.99\columnwidth]{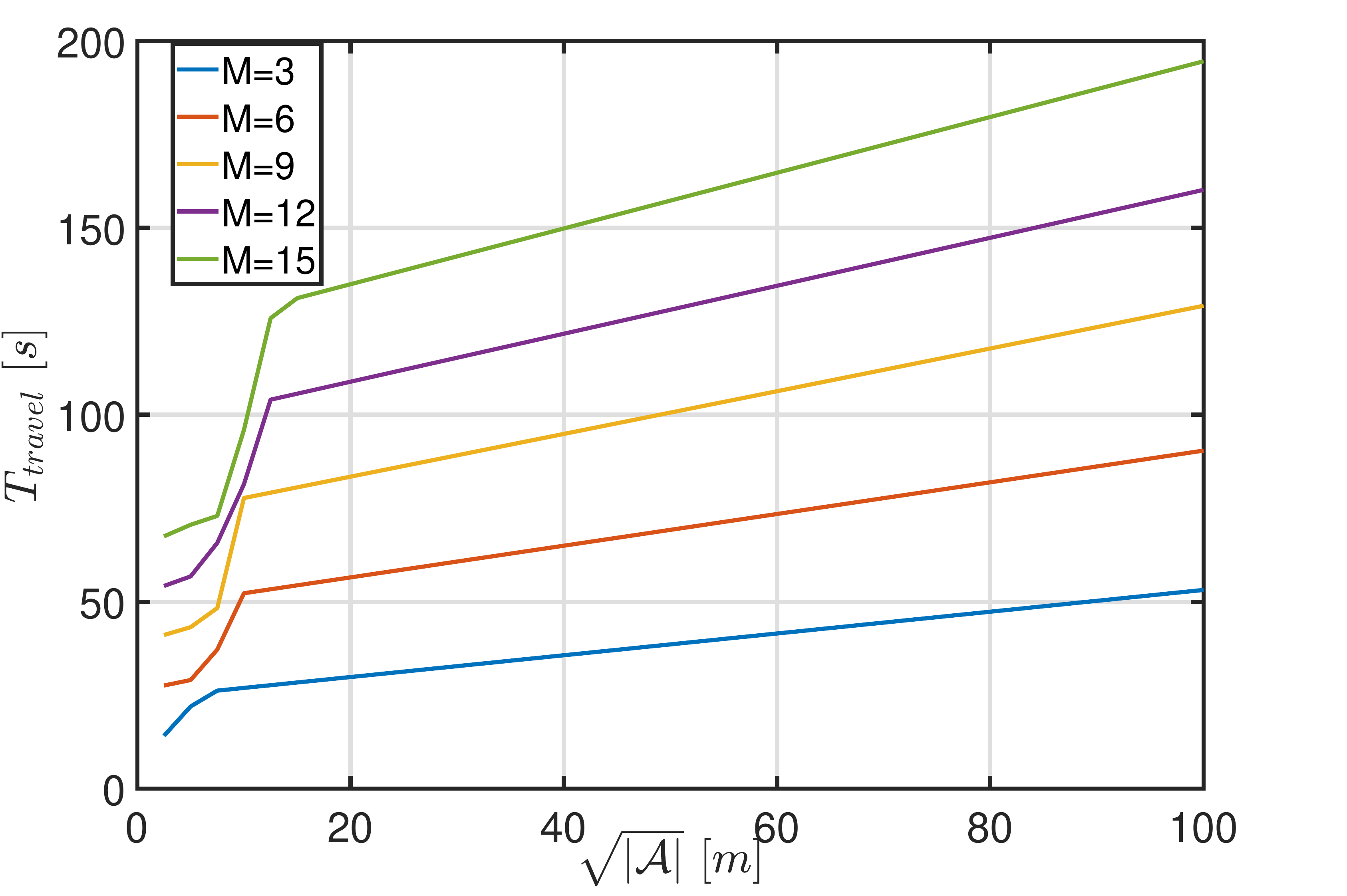}
		\caption{$T_{travel}$ vs. $\sqrt{|\mathcal{A}|}$ for different $M$.}
		\label{fig:T_travel_vs_R}
    \end{subfigure}
    \begin{subfigure}[b]{0.5 \textwidth}
    	\centering
		\includegraphics[width=.99\columnwidth]{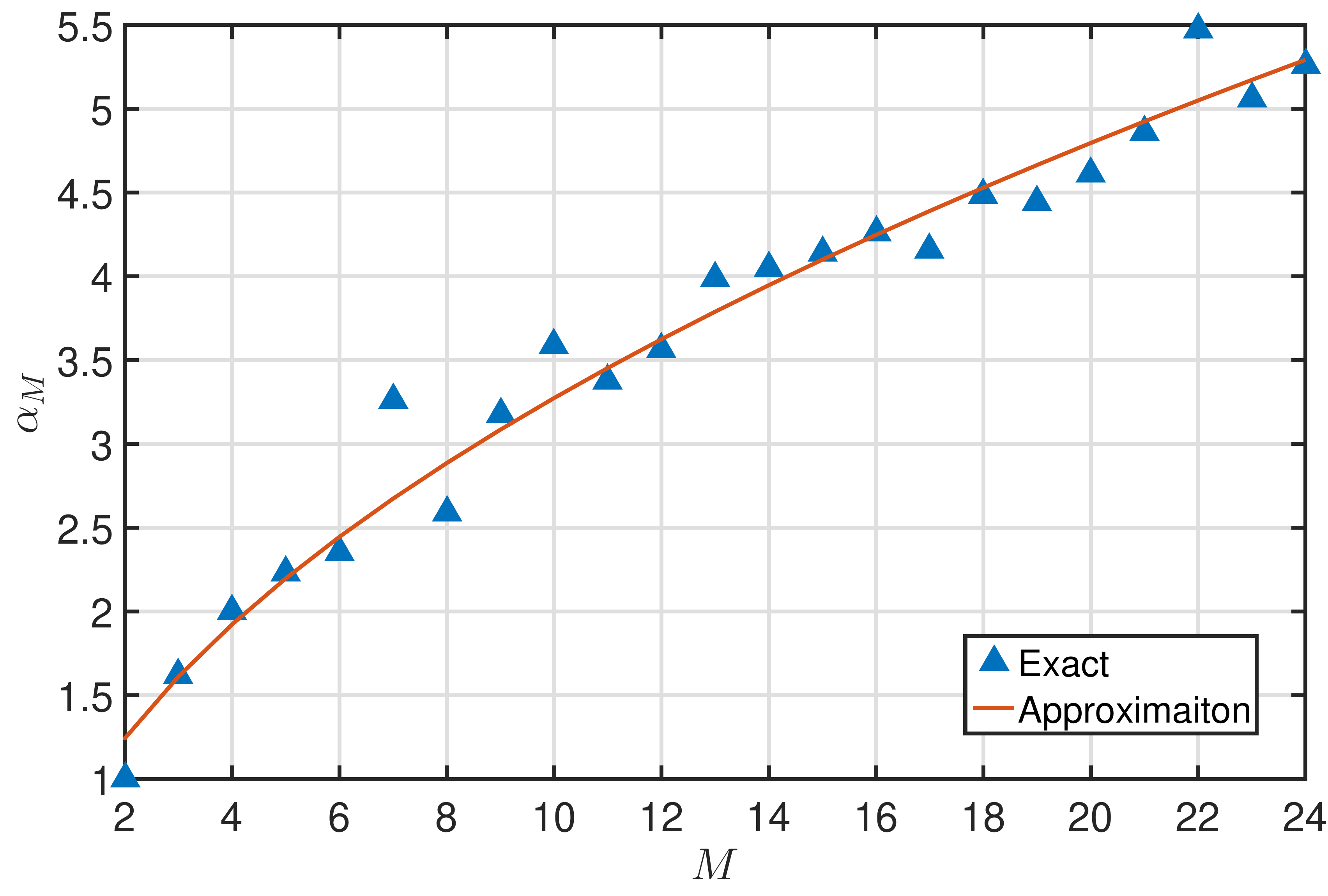}
		\caption{$\alpha_M$ vs. $M$.}
		\label{alpha}
    \end{subfigure}
    \caption{Impact of $\sqrt{|\mathcal{A}|}$ on $T_{travel}$ and traveling distance against $M$ for unit field. }
    \label{fig:travel}
\end{figure}

The traveling time is a function of the number of hovering areas $M$ and the area of the IoT network. As the area gets larger, more time is needed to travel from one hovering location to another. Fig. \ref{fig:T_travel_vs_R} shows $T_{travel}$ against the side length of a squared field for different $M$. For small size of networks, $T_{travel}$ rapidly increases because of the fact that the UAV cannot fully exploit its agility, that is, the traveling distance between HLs is not long enough to reach its maximum speed. On the other hand, $T_{travel}$ starts increasing linearly with $\sqrt{|\mathcal{A}|}$ once the network area provide long hop ranges to exploit the drone agility. Indeed, this was our main motivation to come up with a useful approximation of traveling distance which is applicable for any size of the network area. Fig. \ref{alpha} shows $\alpha_M$ against $M$ as given in Table \ref{table:1}. Curve fitting techniques are utilized to approximate $\alpha_M$ as $\tilde{\alpha}_M = \sqrt{1.35M}-.4$ with an approximation error of $\frac{||\alpha_M - \tilde{\alpha}_M||_2}{||\alpha_M||_2} = 5.5\%$ as shown in Fig. \ref{alpha}. We should note that the approximation error is mainly driven by the $\alpha_M$ outliers which is a result of uncontrollable peculiarity of the coverage problem for integer-valued $M$. Instead of solving the coverage problem and TSP for higher values of $M$, a low-complexity closed-form solution can be directly obtained by approximating the traveling time based on $\tilde{\alpha}_M$.

\begin{figure}[t]
    \centering
    \begin{subfigure}[b]{0.5 \textwidth}
    	\centering
                \includegraphics[width=.99\columnwidth]{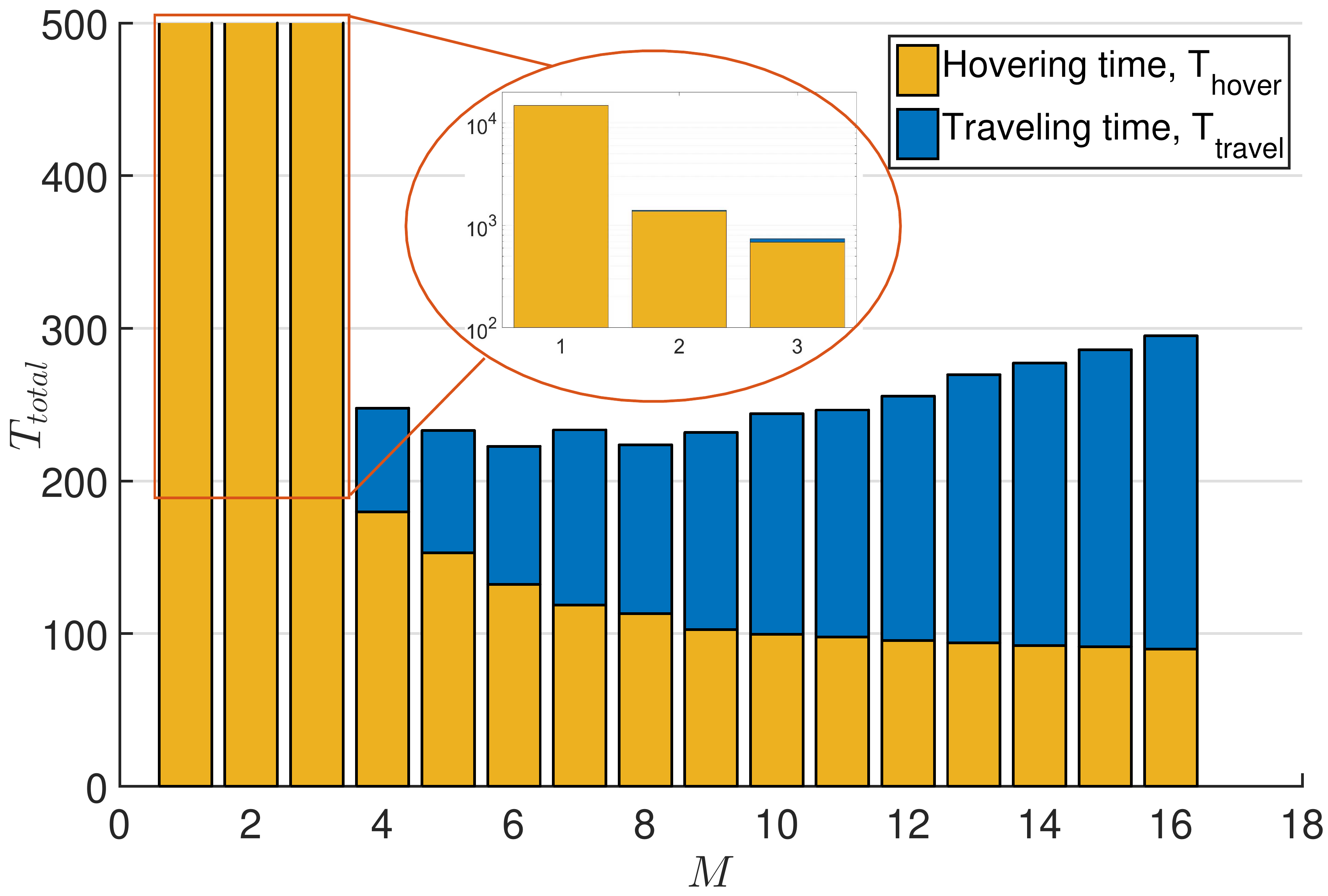}
                \caption{Data aggregation ($\zeta=250$).}
                \label{fig:optimal_M1}
    \end{subfigure}
    \begin{subfigure}[b]{0.5 \textwidth}
    	\centering
                \includegraphics[width=.99\columnwidth]{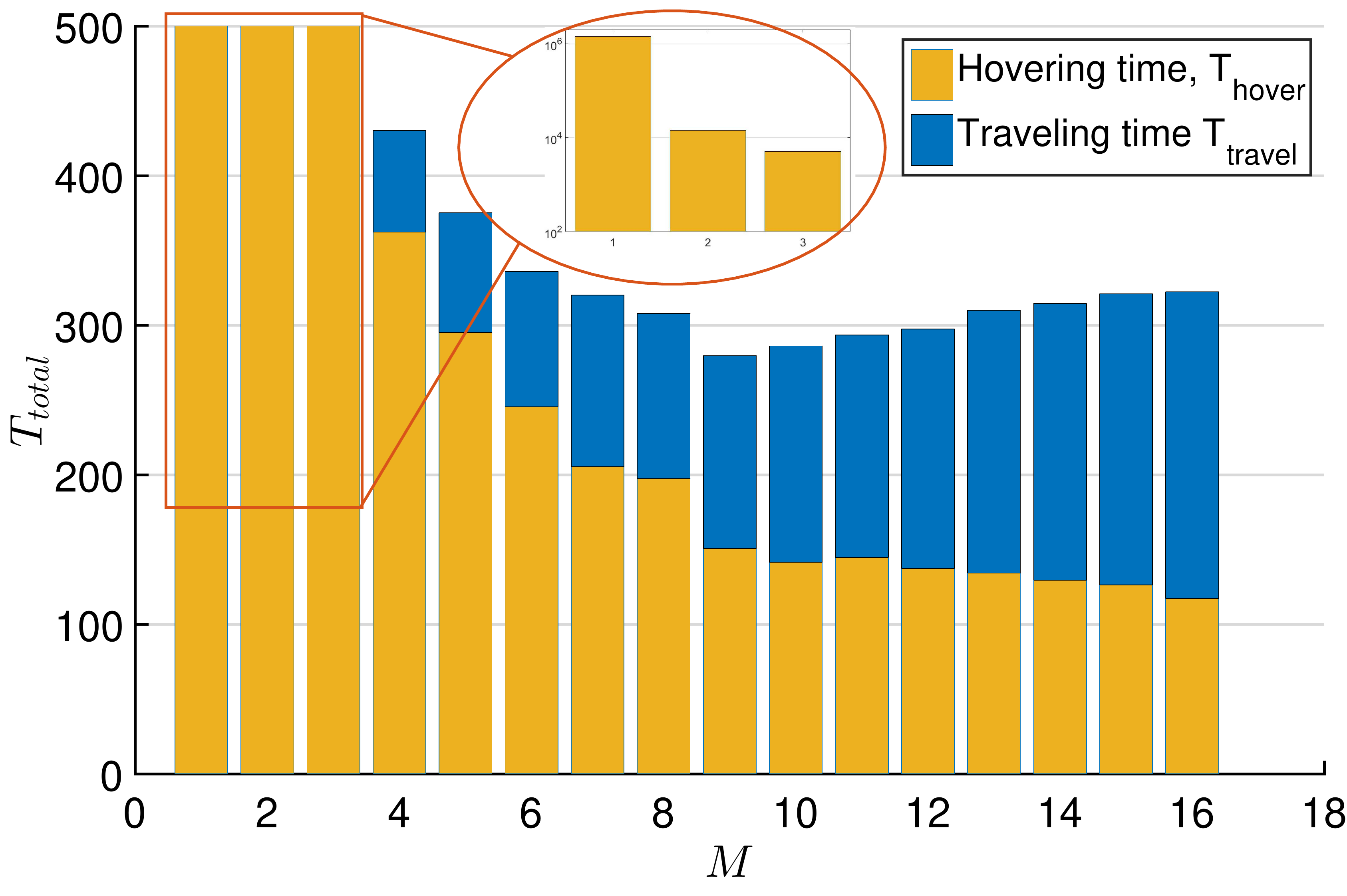}
                \caption{Field estimation ($\delta=0.2$).}
                \label{fig:optimal_M2}
    \end{subfigure}
    \caption{$T_{total}$ vs. $M$.}
    \label{fig:optimal}
\end{figure}
%
%

The impact of $M$ on $T_{total}$, $T_{hover}^\mu$, and $T_{travel}$ for data aggregation mission are demonstrated in Fig~\ref{fig:optimal_M1}. As explained before, the hovering time monotonically decreases with the number of HLs, $M$. Interestingly, the drop is considerable for small values of $M$, then it saddles on a gradual level after $M \geq 4$. In particular, the significantly sharp reduction at $M=4$ can be understood by crosschecking Table \ref{table:2} and Fig. \ref{fig:P_success_T_hover_vs_h}. On the other hand, traveling time increase with $M$ starts playing an important role especially after $M \geq 4$. The total traveling time is determined based on $T_{hover}^\mu$, and $T_{travel}$ and is shown to achieve its minimum at $M=6$ with $T_{total}=223 s$. We must note that the optimal point of $T_{total}$ is subject to different system parameters such as $\zeta$, $\beta$, $S$, $B$, etc.

%
Fig. \ref{fig:optimal_M2} shows the total traveling and hovering time for field estimation against $M$. While data aggregation case reaches the minimal $T_{total}$ at $M=6$, field estimation achieves that at $M=9$. Although traveling time is the same for both missions, more hovering locations is desirable for field estimation mission to satisfy the MSE constraint as previously explained in details. 

\begin{figure}[t]
	\centering
	\begin{subfigure}[b]{0.5 \textwidth}
		\centering
		\includegraphics[width=.99\columnwidth]{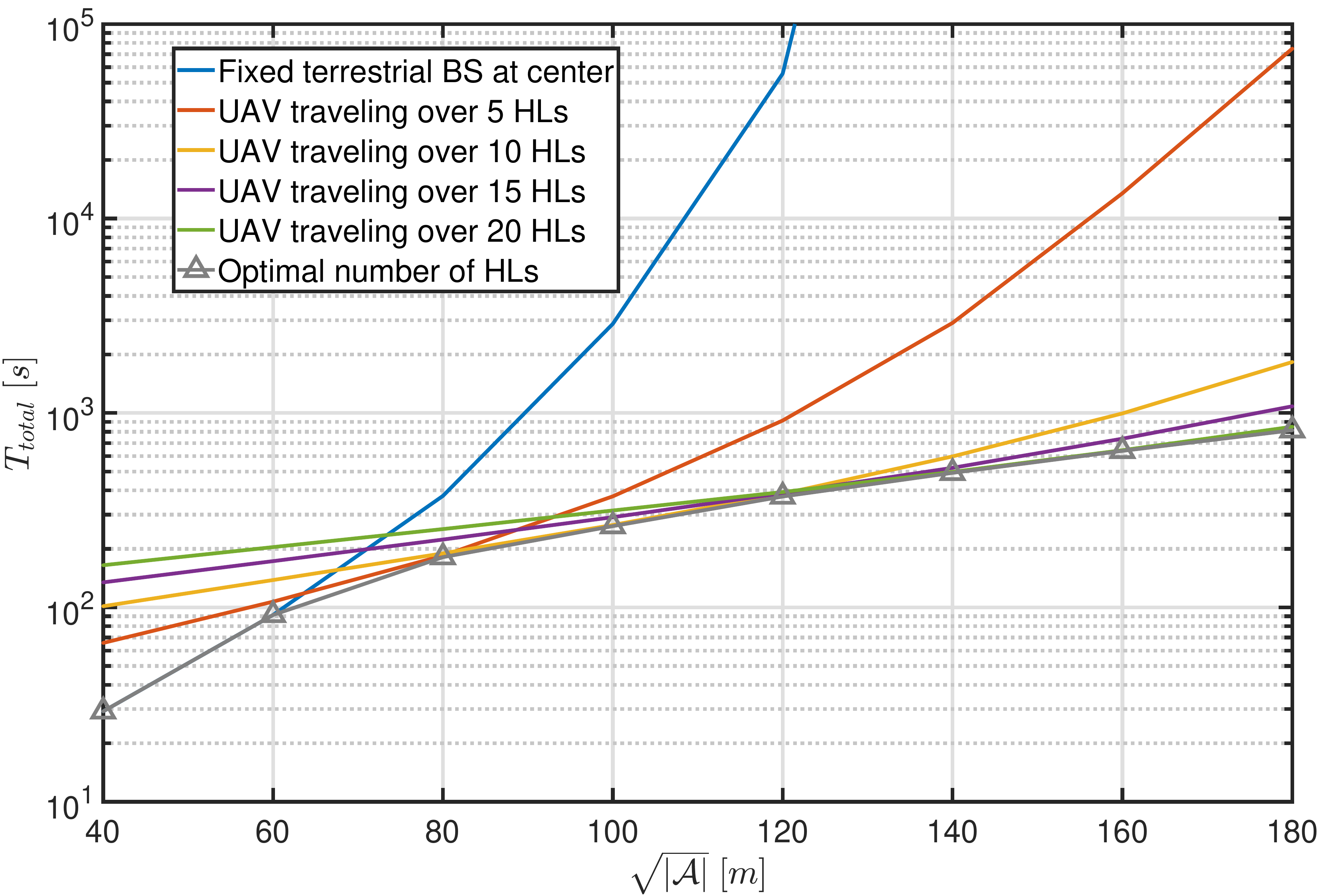}
		\caption{The role of $\sqrt{|\mathcal{A}|}$ on the optimum number of HLs.}
		\label{no_UAV0}
	\end{subfigure}
	\begin{subfigure}[b]{0.5 \textwidth}
		\centering
		\includegraphics[width=.99\columnwidth]{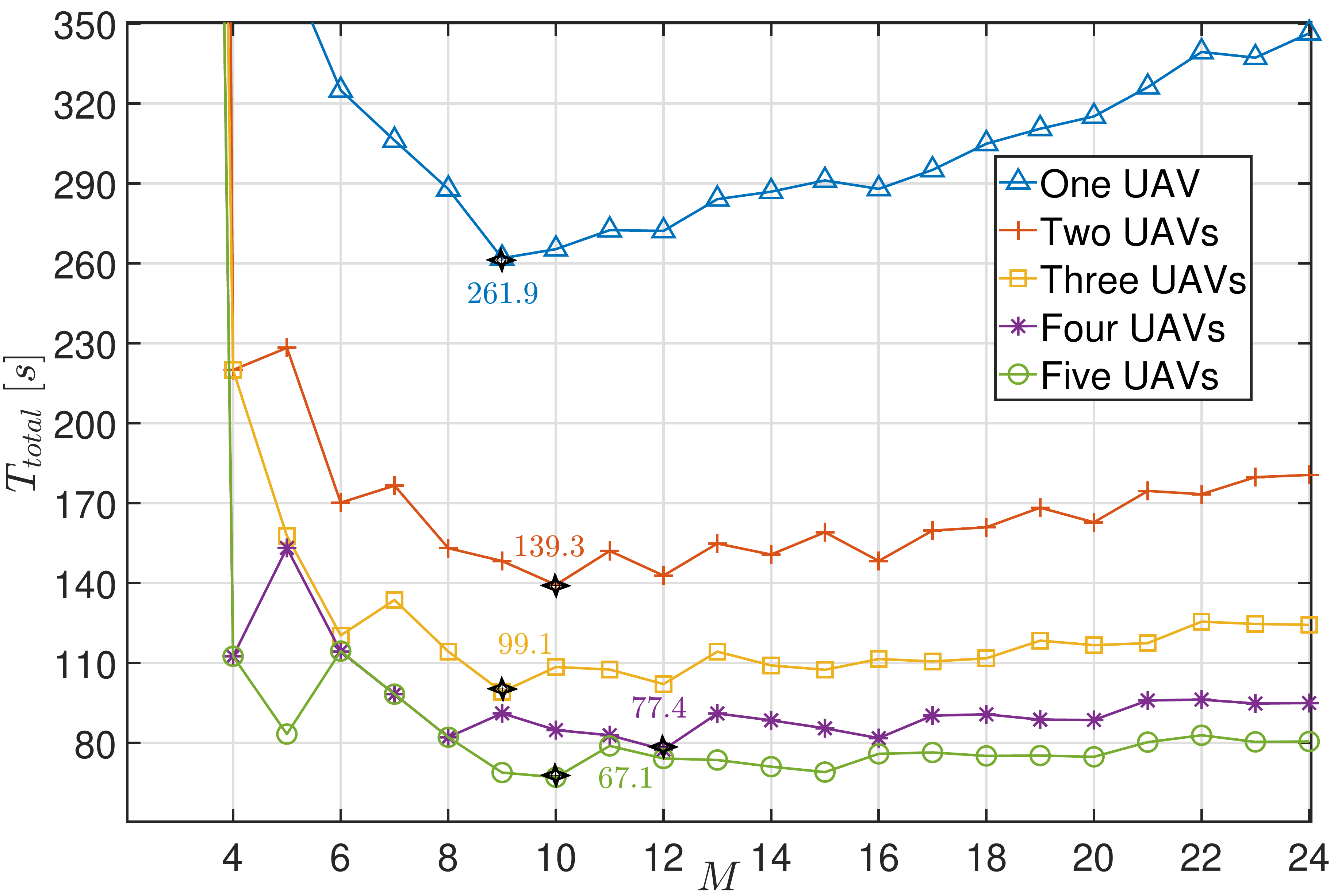}
		\caption{$T_{travel}$ for multi-UAVs with CDS at the center.}
		\label{multi_UAV_total_time}
	\end{subfigure}
	\caption{Optimum number of HLs vs $\sqrt{|\mathcal{A}|}$ and $T_{total}$ for multi-UAVs for field estimation. }
	\label{no_UAV_Multi_UAV}
\end{figure}

The size of the area of interest plays a crucial role in optimum number of HLs and coverage area at each HL. In order to demonstrate the benefits of using UAVs and efficiency of proposed optimization approach, we depict total mission time with respect to increasing field area. As shown in Fig. \ref{no_UAV0}, the optimal number of HLs increases with the field area, which is always captured by the proposed solution. In the conventional data aggregation or field estimation approach, traveling time is not a matter of concern as there is no UAV to deploy. However, to obtain the same UAV benefits (e.g., extended network size), it is necessary to deploy BSs as much as the HLs, which may be costly or even impossible in inaccessible geographical areas. In Fig \ref{multi_UAV_total_time}, we show the total mission time for different number of UAVs against the number of HLs. The UAVs start and end the mission at a CDS placed at the center. Notice that total mission time is constantly reduced by employing more UAVs, which can be further improved by allowing additional CDSs. Also notice that the optimal number of HLs is practically divisible by the number of UAVs. This is indeed expected since every UAV covers almost equal number of HLs and therefore all UAVs have comparable mission times.

\subsection{Time Complexity Analysis}
We finally present complexity analysis of the proposed solution	before concluding the paper. For each $M$ we find $\Delta_M$ and $\alpha_M$ by applying the coverage problem and TSP to construct Table 1. The employed coverage problem solution uses a modified Newton-Raphson method which has a time complexity of $\mathcal{O} \left( \log n F(n)\right)$  where $F(n)$ is the cost of calculating the root with $n$-digit precision \cite{heath2018scientific}. Based on a dynamic programming based approach, the TSP can be solved by Held–Karp algorithm in the order of $\mathcal{O}\left( 2^M \sqrt{M}\right)$. However, we must note that the proposed solution calculates $\Delta_M$ and $\alpha_M$ to be referred as an offline table which is applicable to any field area. Hence, we use (8) to find $T_{travel}$ for any field size and UAV agility parameters with a negligible computational complexity.  
	
	For the data aggregation problem, $\vect{P_1}$, equations (12) and (17) are solved to obtain $T_{hover}$. We optimize $a$ and $\beta$ numerically. For the field estimation mission, $\vect{P_2}$, the probability of successful transmission at the edge is found by solving (25). Then $J_\mu^*$ is obtained by solving (31). Finally, $T_{hover}$ is obtained as in (32). Again, $R_{mse}$, $a$, and $\beta$ are optimized numerically. Employed numerical solution exploits bi-section method with $\mathcal{O} (\log_2 \left(\frac{\epsilon_0}{\epsilon}\right))$ where $\epsilon_0$ is the initial bracket size (e.g., $\epsilon_0=1$ for a probabilistic variable) and $\epsilon$ is the required error tolerance, e.g., $10^{-6}$. We iteratively calculate total mission time for each $M$ starting from $M=1$ until it stops decreasing. Hence, the time complexity of the proposed solution can be given by $\mathcal{O}\left( M^\star \log_2 \left(\frac{\epsilon_0}{\epsilon}\right)  \right)$ where $M^\star$ is the number of HLs with minimum total mission time.		

\section{Conclusion}\label{sec:conclusion}

This paper investigates aerial data aggregation and field estimation missions in IoT networks by integrating stochastic geometry, graph theory, signal processing, and optimization theory. To accomplish these tasks, the field of interest is divided into several subregions over which the UAV hovers to collect necessary samples from the underlying nodes. Accordingly, we formulate and solve two optimization problems to minimize total mission duration. While the former requires the collection of a prescribed average number of samples from the field, the latter ensures for a given a field spatial correlation model that the average mean-squared estimation error of the field value is no more than a predetermined threshold at any point. Minimal mission duration is obtained by optimizing the number of subregions, the area of each subregion, the hovering locations, the hovering time at each location, and the trajectory traversed between hovering locations. The proposed formulation is {\em NP-hard mixed integer} problem, and hence, a decoupled heuristic solution is proposed by augmenting the hovering and traveling time dilemma. In particular, a closed-form traveling time approximation is proposed to avoid computational complexity of integer part of the problem, the solution of which is quite time-consuming even for heuristic solutions. For a given number of HLs, hovering time is formulated by means of the stochastic geometry. Optimal transmission probability and SINR threshold are also obtained analytically and numerically for the adopted slotted ALOHA scheme, respectively. The results show that there exists an optimal number of subregions that balance the tradeoff between hovering and traveling times such that the total time for collecting the required samples is minimized.

\begin{appendices}
\section{Proof of Theorem 1}
\label{app:success}
We start by deriving the Laplace transform of the normalized interference plus noise power distribution,
\begin{align*}
	&\mathcal{L}_I(s) \overset{(a)}{=} \mathbb{E}_I [ e^{-sI}]  \overset{(b)}{=} \mathbb{E}_D \mathbb{E}_G \Big[ e^{ -s\sigma_n^2/P} \prod_{\bf{x} \in \tilde{\Psi}_\mu \setminus \bf{x}_n} e^{-s G_x D_x^{-\eta}}\Big] \\	
	&\overset{(c)}{=} e^{ -s\sigma_n^2/P} \, \mathbb{E}_D          \Big[ \prod_{\bf{x} \in \tilde{\Psi}_\mu \setminus \bf{x}_n}  \Big(  1+\dfrac{s D_x^{-\eta}}{m}   \Big)^{-m} \Big] \\
	&\overset{(d)}{=}  e^{ -s\sigma_n^2/P} \exp \Bigg( -2 a \pi \lambda \int_{h}^{d} \Big( 1- \big(1+\frac{s r^{-\eta}}{m} \big) ^{-m} \Big) r dr    \Bigg),\\
\end{align*} 
where $(a)$ follows from the definition of the Laplace transform. $(b)$ follows from the independence between the PDFs $f_D(r_x)$ and $f_G(g)$. In $(c)$, since the channel gains of different nodes are i.i.d., the product of expectations $\prod \mathbb{E}_{G_x}(\cdot)$ is equal to the expectation of the product $\mathbb{E}_{G} \prod (\cdot)$ where $\mathbb{E}_{G_x}(\cdot)$ follows from the moment generating function (MGF) of gamma distribution. And $(d)$ follows from the probability generating functional (PGFL) of PPP process and the distance distribution as given in \eqref{eq:distance_tx}.   
The coverage probability is derived in terms of $\mathcal{L}_I(s)$ as follows,
\begin{align*}
	&P_{\mu}^s =  \mathbb{P}(\bigcup_{\bf{x}_n \in \tilde{\Psi}_\mu} \text{SINR}_\mu^n >\beta) \\
	&\overset{(a)}{=}  \mathbb{E}_{D_n} \mathbb{E}_I \Big[ \sum_{\bf{x}_n \in \tilde{\Psi}_\mu} \mathbb{P} \big( G_n > \beta D_n^\eta I|D_n,I \big)  \Big] \\
	&\overset{(b)}{=}  \mathbb{E}_{D_n} \mathbb{E}_I \Big[  \sum_{\bf{x}_n \in \tilde{\Psi}_\mu} \dfrac{\Gamma(m,m\beta r^\eta I)}{\Gamma(m)} \Big| D_n \Big] \\
	&\overset{(c)}{=}  \mathbb{E}_{D_n} \mathbb{E}_I \Big[ \sum_{\bf{x}_n \in \tilde{\Psi}_\mu} \sum_{k=0}^{m-1} \dfrac{ (m\beta r^\eta I)^k}{k!} e^{-m\beta r^\eta I} \Big| D_n \Big]\\
	&\overset{(d)}{=} 2 a \pi \lambda \int_{h}^{d} \sum_{k=0}^{m-1}  \dfrac{(-m\beta r^\eta)^k}{k!} \Big[ \dfrac{\partial^k}{\partial s^k} \mathcal{L}_I(s) \Big]_{s= m\beta r^\eta} {r} \, dr, 
\end{align*}
where in $(a)$ the union is equivalent to summation since the events are mutually exclusive, \cite{Dhillon2012}. $(b)$ follows from the CCDF of $G_n$ and that $\mathbb{P}(\text{SINR}_\mu^n >\beta) $ are i.i.d over $n$. $(c)$ follows from the incomplete gamma function definition for $m \in \mathbb{Z}^+$. $(d)$ follows from Campbell Mecke Theorem\cite{Macke2013}.

\section{Proof of proposition 1} \label{proposition}
In order to obtain the upper bound on the MSE of an edge point, $s_e$, we consider two cases: Neglecting the correlation of points whose distance to $s_e$ is more than $R_{mse}$, we first focus on the occurrence of no observation is received from points within $\mathcal{A}_{int}$. In this case, upper bound on the MSE is set to $E_e \leq \sigma^2$ by counting on the observations obtained from outside of $\mathcal{A}_{int}$. Second, we consider the case where an observation is  received from a point within $\mathcal{A}_{int}$. Denoting the distance between $s_e$ and such a point by $x$, the exponential covariance model sets upper bound on the MSE as $E_e \leq \sigma^2 - \dfrac{\exp(-2x/b)}{\sigma^2} \leq \sigma^2 - \dfrac{\exp(-2R_{mse}/b)}{\sigma^2}$. Since the first and second cases have a probability of $P_{e}^{n-s}(R_{mse},J_\mu)$ and $1- P_{e}^{n-s}(R_{mse},J_\mu)$, respectively, the expected MSE upper bound can be obtained as in \eqref{eq:mse_inequality}.

\section{proof of Theorem 2}\label{app:p_success_e}
The integration of a function $f(w)$ over the area of intersection between two disks $b(0,R)$ and $b(R, R_{mse})$ is given in \cite{Niyato2017} as,
\begin{align}\label{eq:int_two_disks}
\int_{0}^{R} w f(w) \theta(w) dw,
\end{align}
where $w$ represents the distance from the center of the disk $b(0,R)$ and $\theta$ is as in \eqref{eq:theta}. In \eqref{eq:P_cov}, $P_{\mu}^s$ is given by integrating the coverage probability as a function of the distance between the UAV and a transmitting sensor, $r$, over the area of the disk $b(0,R)$). i.e,
\begin{align}
P_{\mu}^s = \int_{0}^{2\pi} \int_{h}^{d} P_{\mu}^*(r) dr d\theta.
\end{align}
Hence,
\begin{align}
P_{\mu}^*(r) = \dfrac{1}{\pi R^2}\sum_{k=0}^{m_0-1} n \dfrac{(-m_0\beta r^a)^k}{k!}  \Big[ \dfrac{\partial^k}{\partial s^k} \mathcal{L}_I(s) \Big]_{s= m_0\beta r^a} .
\end{align}
By substituting $r = \sqrt{w^2+h^2}$ and $f(w) = P_{\mu}^*(r)$ in \eqref{eq:int_two_disks}, the coverage probability over the area of intersection between the two disks $b(0,R)$ and $b(R, R_{mse})$ is expressed as,
\begin{align}
P_{e}^s =  \int_{h}^{d} r P_{\mu}^*(r) \theta(\sqrt{r^2-h^2}) dr,
\end{align}
which proves the Theorem. 

\end{appendices}



\bibliographystyle{IEEEtran}
\bibliography{IEEEabrv,Ref}

\begin{thebibliography}{10}
\providecommand{\url}[1]{#1}
\csname url@samestyle\endcsname
\providecommand{\newblock}{\relax}
\providecommand{\bibinfo}[2]{#2}
\providecommand{\BIBentrySTDinterwordspacing}{\spaceskip=0pt\relax}
\providecommand{\BIBentryALTinterwordstretchfactor}{4}
\providecommand{\BIBentryALTinterwordspacing}{\spaceskip=\fontdimen2\font plus
\BIBentryALTinterwordstretchfactor\fontdimen3\font minus
  \fontdimen4\font\relax}
\providecommand{\BIBforeignlanguage}[2]{{%
\expandafter\ifx\csname l@#1\endcsname\relax
\typeout{** WARNING: IEEEtran.bst: No hyphenation pattern has been}%
\typeout{** loaded for the language `#1'. Using the pattern for}%
\typeout{** the default language instead.}%
\else
\language=\csname l@#1\endcsname
\fi
#2}}
\providecommand{\BIBdecl}{\relax}
\BIBdecl

\bibitem{Iot2015}
A.~Al-Fuqaha, M.~Guizani, M.~Mohammadi, M.~Aledhari, and M.~Ayyash, ``Internet
  of things: A survey on enabling technologies, protocols, and applications,''
  \emph{IEEE Communications Surveys Tutorials}, vol.~17, no.~4, pp. 2347--2376,
  Fourthquarter 2015.

\bibitem{Kamal}
J.~N. Al-Karaki and A.~E. Kamal, ``Routing techniques in wireless sensor
  networks: A survey,'' \emph{IEEE Wireless Communications}, vol.~11, no.~6,
  pp. 6--28, Dec 2004.

\bibitem{Zeng2016May}
Y.~Zeng, R.~Zhang, and T.~J. Lim, ``Wireless communications with unmanned
  aerial vehicles: Opportunities and challenges,'' \emph{IEEE Communications
  Magazine}, vol.~54, no.~5, pp. 36--42, May 2016.

\bibitem{LTE_Sky}
B.~V.~D. Bergh, A.~Chiumento, and S.~Pollin, ``{LTE} in the sky: Trading off
  propagation benefits with interference costs for aerial nodes,'' \emph{IEEE
  Communications Magazine}, vol.~54, no.~5, pp. 44--50, May 2016.

\bibitem{Public_UAV}
A.~Merwaday, A.~Tuncer, A.~Kumbhar, and I.~Guvenc, ``Improved throughput
  coverage in natural disasters: Unmanned aerial base stations for
  public-safety communications,'' \emph{IEEE Vehicular Technology Magazine},
  vol.~11, no.~4, pp. 53--60, Dec 2016.

\bibitem{khuwaja2018survey}
A.~A. Khuwaja, Y.~Chen, N.~Zhao, M.~Alouini, and P.~Dobbins, ``A survey of
  channel modeling for {UAV} communications,'' \emph{IEEE Communications
  Surveys Tutorials}, pp. 1--1, 2018.

\bibitem{Sun2015April}
R.~Sun and D.~W. Matolak, ``Initial results for airframe shadowing in {L}- and
  {C}-band air-ground channels,'' in \emph{2015 Integrated Communication,
  Navigation and Surveillance Conference (ICNS)}, April 2015.

\bibitem{Sun2015June}
D.~W. Matolak and R.~Sun, ``Unmanned aircraft systems: Air-ground channel
  characterization for future applications,'' \emph{IEEE Vehicular Technology
  Magazine}, vol.~10, no.~2, pp. 79--85, June 2015.

\bibitem{Akram2014Dec}
A.~Al-Hourani, S.~Kandeepan, and A.~Jamalipour, ``Modeling air-to-ground path
  loss for low altitude platforms in urban environments,'' in \emph{2014 IEEE
  Global Communications Conference}, Dec 2014.

\bibitem{Akram2014Dec2}
A.~Al-Hourani, S.~Kandeepan, and S.~Lardner, ``Optimal {LAP} altitude for
  maximum coverage,'' \emph{IEEE Wireless Communications Letters}, vol.~3,
  no.~6, pp. 569--572, Dec 2014.

\bibitem{Akram2016Dec}
A.~AL-Hourani, S.~Chandrasekharan, G.~Kaandorp, W.~Glenn, A.~Jamalipour, and
  S.~Kandeepan, ``Coverage and rate analysis of aerial base stations,''
  \emph{IEEE Transactions on Aerospace and Electronic Systems}, vol.~52, no.~6,
  pp. 3077--3081, December 2016.

\bibitem{Dhillon}
V.~V. Chetlur and H.~S. Dhillon, ``Downlink coverage analysis for a finite
  3-{D} wireless network of unmanned aerial vehicles,'' \emph{IEEE Transactions
  on Communications}, vol.~65, no.~10, pp. 4543--4558, Oct 2017.

\bibitem{Mozaffari1}
M.~Mozaffari, W.~Saad, M.~Bennis, and M.~Debbah, ``Wireless communication using
  unmanned aerial vehicles ({UAV}s): Optimal transport theory for hover time
  optimization,'' \emph{IEEE Transactions on Wireless Communications}, vol.~16,
  no.~12, pp. 8052--8066, Dec 2017.

\bibitem{Mozaffari2017Nov}
------, ``Mobile unmanned aerial vehicles ({UAV}s) for energy-efficient
  internet of things communications,'' \emph{IEEE Transactions on Wireless
  Communications}, vol.~16, no.~11, pp. 7574--7589, Nov 2017.

\bibitem{Esrafilian2019}
O.~{Esrafilian}, R.~{Gangula}, and D.~{Gesbert}, ``Learning to communicate in
  {UAV}-aided wireless networks: Map-based approaches,'' \emph{IEEE Internet of
  Things Journal}, 2019.

\bibitem{Esrafilian2019SPAWC}
R.~{Gangula}, O.~{Esrafilian}, D.~{Gesbert}, C.~{Roux}, F.~{Kaltenberger}, and
  R.~{Knopp}, ``Flying rebots: First results on an autonomous {UAV}-based {LTE}
  relay using open airinterface,'' in \emph{2018 IEEE 19th International
  Workshop on Signal Processing Advances in Wireless Communications (SPAWC)},
  June 2018.

\bibitem{Mozaffari2016D2D}
M.~{Mozaffari}, W.~{Saad}, M.~{Bennis}, and M.~{Debbah}, ``Unmanned aerial
  vehicle with underlaid device-to-device communications: Performance and
  tradeoffs,'' \emph{IEEE Transactions on Wireless Communications}, vol.~15,
  no.~6, pp. 3949--3963, June 2016.

\bibitem{Mozaffari2019}
M.~{Mozaffari}, A.~{Taleb Zadeh Kasgari}, W.~{Saad}, M.~{Bennis}, and
  M.~{Debbah}, ``Beyond 5{G} with {UAV}s: Foundations of a 3{D} wireless
  cellular network,'' \emph{IEEE Transactions on Wireless Communications},
  vol.~18, no.~1, pp. 357--372, Jan 2019.

\bibitem{MoeWin}
A.~Giorgetti, M.~Lucchi, M.~Chiani, and M.~Z. Win, ``Throughput per pass for
  data aggregation from a wireless sensor network via a {UAV},'' \emph{IEEE
  Transactions on Aerospace and Electronic Systems}, vol.~47, no.~4, pp.
  2610--2626, Oct 2011.

\bibitem{traj_UAV}
R.~Sugihara and R.~K. Gupta, ``Optimal speed control of mobile node for data
  collection in sensor networks,'' \emph{IEEE Transactions on Mobile
  Computing}, vol.~9, no.~1, pp. 127--139, Jan 2010.

\bibitem{wackerly2002mathematical}
D.~Wackerly, W.~Mendenhall, and R.~Scheaffer, \emph{Mathematical Statistics
  with Applications}.\hskip 1em plus 0.5em minus 0.4em\relax Pacific Grove CA,
  USA: Duxbury, 2002.

\bibitem{simon2005digital}
M.~K. Simon and M.-S. Alouini, \emph{Digital Communication over Fading
  Channels}.\hskip 1em plus 0.5em minus 0.4em\relax John Wiley \& Sons, 2005,
  vol.~95.

\bibitem{6}
Q.~{Wu}, Y.~{Zeng}, and R.~{Zhang}, ``Joint trajectory and communication design
  for multi-uav enabled wireless networks,'' \emph{IEEE Transactions on
  Wireless Communications}, vol.~17, no.~3, pp. 2109--2121, March 2018.

\bibitem{7}
F.~{Cheng}, S.~{Zhang}, Z.~{Li}, Y.~{Chen}, N.~{Zhao}, F.~R. {Yu}, and V.~C.~M.
  {Leung}, ``Uav trajectory optimization for data offloading at the edge of
  multiple cells,'' \emph{IEEE Transactions on Vehicular Technology}, vol.~67,
  no.~7, pp. 6732--6736, July 2018.

\bibitem{8}
Y.~{Zeng} and R.~{Zhang}, ``Energy-efficient uav communication with trajectory
  optimization,'' \emph{IEEE Transactions on Wireless Communications}, vol.~16,
  no.~6, pp. 3747--3760, June 2017.

\bibitem{Miller1960}
C.~E. Miller, A.~W. Tucker, and R.~A. Zemlin, ``Integer programming formulation
  of traveling salesman problems,'' \emph{J. ACM}, vol.~7, no.~4, pp. 326--329,
  Oct. 1960.

\bibitem{Nurmela2000CoveringAS}
K.~J. Nurmela and P.~R.~J. Östergård, ``Covering a square with up to 30 equal
  circles,'' Helsinki University of Technology, Laboratory for Theoretical
  Computer Science. Research report, 2000.

\bibitem{Stoyan2010}
Y.~G. Stoyan and V.~M. Patsuk, ``Covering a compact polygonal set by identical
  circles,'' \emph{Computational Optimization and Applications}, vol.~46,
  no.~1, pp. 75--92, May 2010.

\bibitem{TSP1997}
M.~Dorigo and L.~M. Gambardella, ``Ant colony system: a cooperative learning
  approach to the traveling salesman problem,'' \emph{IEEE Transactions on
  Evolutionary Computation}, vol.~1, no.~1, pp. 53--66, Apr 1997.

\bibitem{Kirk2014TSP}
J.~Kirk, ``Traveling salesman problem-genetic algorithm, version 1.3 (2014),''
  \url{https://www.mathworks.com/matlabcentral/fileexchange/13680-traveling-salesman-problem-genetic-algorithm},
  MATLAB Central File Exchange, Retrieved, Jan. 2018.

\bibitem{Bushnaq2018Aerial}
O.~M. Bushnaq, A.~Celik, H.~ElSawy, M.-S. Alouini, and T.~Y. Al-Naffouri,
  ``Aerial data aggregation in {IoT} networks: Hovering \& traveling time
  dilemma,'' in \emph{2018 IEEE Global Communications Conference}, vol.
  Abu-Dhabi, UAE, Dec 2018.

\bibitem{Esrafilian2018}
O.~{Esrafilian} and D.~{Gesbert}, ``Simultaneous user association and placement
  in multi-{UAV} enabled wireless networks,'' in \emph{WSA 2018; 22nd
  International ITG Workshop on Smart Antennas}, March 2018.

\bibitem{Wu2018}
Q.~Wu and R.~Zhang, ``Common throughput maximization in {UAV}-enabled ofdma
  systems with delay consideration,'' \emph{IEEE Transactions on
  Communications}, vol.~66, pp. 6614--6627, 2018.

\bibitem{Li2017}
J.~{Li} and Y.~{Han}, ``Optimal resource allocation for packet delay
  minimization in multi-layer {UAV} networks,'' \emph{IEEE Communications
  Letters}, vol.~21, no.~3, pp. 580--583, March 2017.

\bibitem{Zhang2018Analysis}
S.~{Zhang} and J.~{Liu}, ``Analysis and optimization of multiple unmanned
  aerial vehicle-assisted communications in post-disaster areas,'' \emph{IEEE
  Trans. Vehicular Technol.}, vol.~67, no.~12, pp. 12\,049--12\,060, Dec. 2018.

\bibitem{Chen2018Distributed}
J.~{Chen}, Q.~{Wu}, Y.~{Xu}, Y.~{Zhang}, and Y.~{Yang}, ``Distributed
  demand-aware channel-slot selection for multi-uav networks: A game-theoretic
  learning approach,'' \emph{IEEE Access}, vol.~6, pp. 14\,799--14\,811, 2018.

\bibitem{M_TSP}
K.~V. Narasimha, E.~Kivelevitch, B.~Sharma, and M.~Kumar, ``An ant colony
  optimization technique for solving min-max multi-depot vehicle routing
  problem,'' \emph{Swarm and Evolutionary Computation}, vol.~13, pp. 63 -- 73,
  2013.

\bibitem{Elad2011MTSP}
E.~Kivelevitch, ``{MDMTSPV} {GA} multiple depot multiple traveling salesmen
  problem solved by genetic algorithm, version 1.0.0.0 (2011),''
  \url{https://www.mathworks.com/matlabcentral/fileexchange/31814-mdmtspv\_ga-multiple-depot-multiple-traveling-salesmen-problem-solved-by-genetic-algorithm},
  MATLAB Central File Exchange, Retrieved, Jun. 2011.

\bibitem{Directional_Antenna}
J.~{Lyu}, Y.~{Zeng}, R.~{Zhang}, and T.~J. {Lim}, ``Placement optimization of
  {UAV}-mounted mobile base stations,'' \emph{IEEE Communications Letters},
  vol.~21, no.~3, pp. 604--607, March 2017.

\bibitem{Dhillon2012}
H.~S. Dhillon, R.~K. Ganti, F.~Baccelli, and J.~G. Andrews, ``Modeling and
  analysis of k-tier downlink heterogeneous cellular networks,'' \emph{IEEE
  Journal on Selected Areas in Communications}, vol.~30, no.~3, pp. 550--560,
  April 2012.

\bibitem{Marc2018}
Y.~Yan and M.~G. Genton, ``Gaussian likelihood inference on data from
  trans-{G}aussian random fields with {M}atérn covariance function,''
  \emph{Environmetrics}, vol.~29, no. 5-6, p. e2458, 2018.

\bibitem{11}
H.~Zhang and A.~El-Shaarawi, ``On spatial skew-gaussian processes and
  applications,'' \emph{Environmetrics}, vol.~21, no.~1, pp. 33--47, 2010.

\bibitem{12}
L.~Benoit, D.~Allard, and G.~Mariethoz, ``Stochastic rainfall modeling at
  sub-kilometer scale,'' \emph{Water Resources Research}, vol.~54, no.~6, pp.
  4108--4130, 2018.

\bibitem{13}
R.~Hewer, P.~Friederichs, A.~Hense, and M.~Schlather, ``A matérn-based
  multivariate gaussian random process for a consistent model of the horizontal
  wind components and related variables,'' \emph{Journal of the Atmospheric
  Sciences}, vol.~74, no.~11, pp. 3833--3845, 2017.

\bibitem{14}
K.~Feldmann, M.~Scheuerer, and T.~L. Thorarinsdottir, ``Spatial postprocessing
  of ensemble forecasts for temperature using nonhomogeneous gaussian
  regression,'' \emph{Monthly Weather Review}, vol. 143, no.~3, pp. 955--971,
  2015.

\bibitem{Marc2014}
Y.~Sun, K.~P. Bowman, M.~G. Genton, and A.~Tokay, ``A {M}atérn model of the
  spatial covariance structure of point rain rates,'' 2014.

\bibitem{Peter2006}
P.~Guttorp and T.~Gneiting, ``{Studies in the history of probability and
  statistics {XLIX} On the {M}atérn correlation family},'' \emph{Biometrika},
  vol.~93, no.~4, pp. 989--995, 12 2006.

\bibitem{schabenberger2005statistical}
O.~Schabenberger and C.~A. Gotway, \emph{Statistical methods for spatial data
  analysis}.\hskip 1em plus 0.5em minus 0.4em\relax Chapman \& Hall/CRC, 2005.

\bibitem{Atif2018EH}
A.~Bakytbekov, T.~Q. Nguyen, C.~Huynh, K.~N. Salama, and A.~Shamim, ``Fully
  printed 3{D} cube-shaped multiband fractal rectenna for ambient {RF} energy
  harvesting,'' \emph{Nano Energy}, vol.~53, pp. 587 -- 595, 2018.

\bibitem{Roberts1975}
L.~G. Roberts, ``Aloha packet system with and without slots and capture,''
  \emph{SIGCOMM Comput. Commun. Rev.}, vol.~5, no.~2, pp. 28--42, Apr. 1975.

\bibitem{heath2018scientific}
M.~T. Heath, \emph{Scientific computing: an introductory survey}.\hskip 1em
  plus 0.5em minus 0.4em\relax SIAM, 2018.

\bibitem{Macke2013}
S.~N. Chiu, D.~Stoyan, W.~S. Kendall, and J.~Mecke, \emph{Stochastic Geometry
  and Its Applications}.\hskip 1em plus 0.5em minus 0.4em\relax John Wiley \&
  Sons, 2013.

\bibitem{Niyato2017}
I.~Flint, H.~B. Kong, N.~Privault, P.~Wang, and D.~Niyato, ``Analysis of
  heterogeneous wireless networks using {P}oisson hard-core hole process,''
  \emph{IEEE Transactions on Wireless Communications}, vol.~16, no.~11, pp.
  7152--7167, Nov 2017.

\end{thebibliography}


\begin{IEEEbiography}[{\includegraphics[trim={0cm 2cm  0cm 0cm},clip,  width=1.1in]{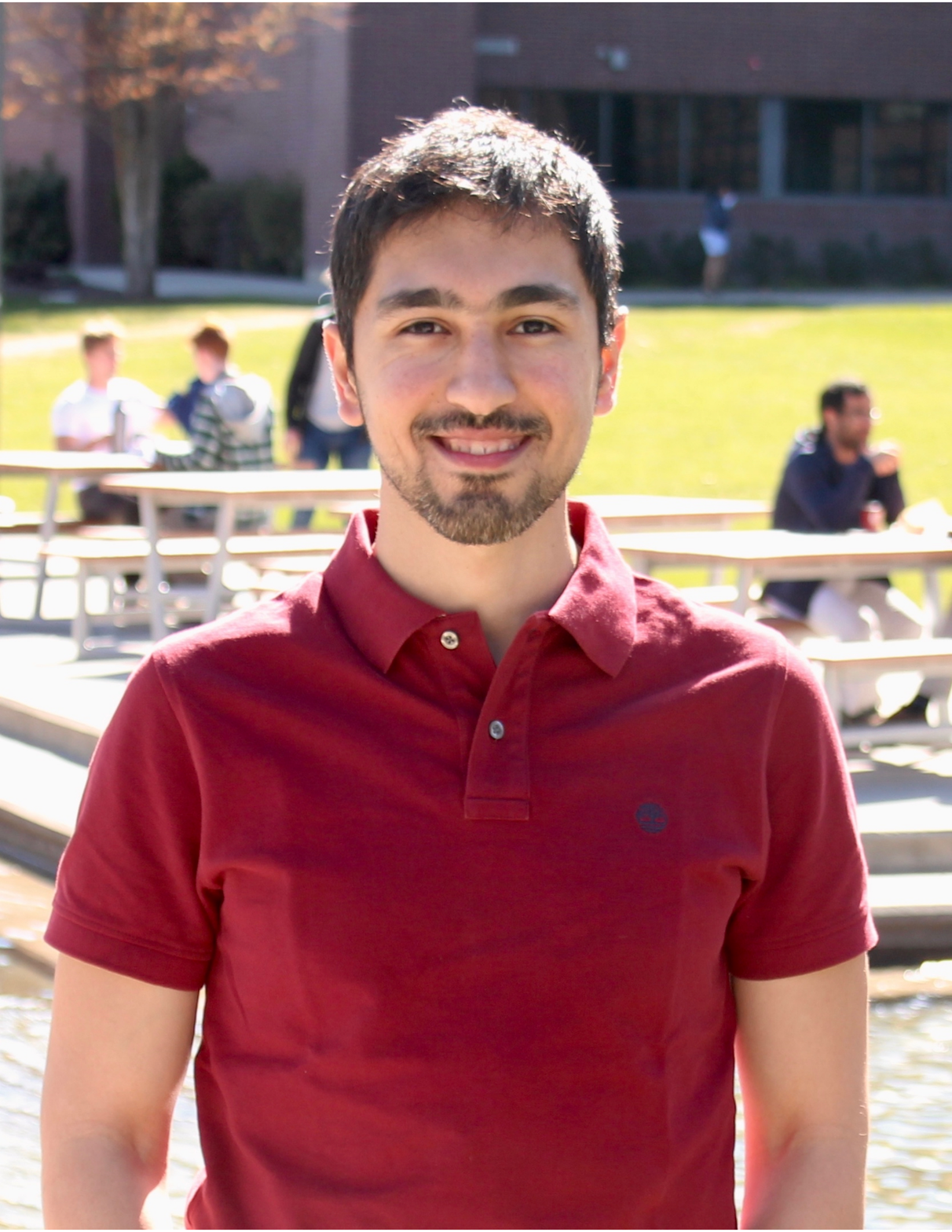}}]{Osama M. Bushnaq}(S'17) received the B.S. degree in communication engineering from Princess Sumaya University for Science and Technology (PSUT), Amman, Jordan in 2012, the M.Eng. degree in electrical engineering from University of New Brunswick (UNB), Fredericton, NB, Canada in 2014. He is currently a Ph.D. candidate at King Abdullah University of Science and Technology (KAUST). He was a visiting researcher at Delft University of Technology (TU Delft), Netherlands in 2016 and University of British Columbia (UBC), Canada in 2019. His current research interests include statistical signal processing, IoT networks, wireless communications, discrete optimization and UAV wireless networks.
\end{IEEEbiography}

\begin{IEEEbiography}[{\includegraphics[width=1.1in,height=1.25in]{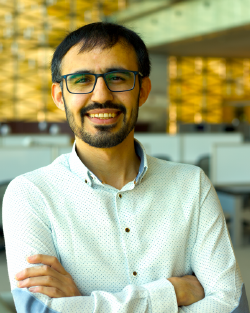}}]{Abdulkadir \c Celik}
	(S'14-M'16) received the B.S. degree in electrical-electronics engineering from Sel\c cuk University, Konya, Turkey in 2009, the M.S. degree in electrical engineering in 2013, the M.S. degree in computer engineering in 2015, and the Ph.D. degree in co-majors of electrical engineering and computer engineering in 2016, all from Iowa State University, Ames, IA, USA. He is currently a postdoctoral research fellow at Communication Theory Laboratory of King Abdullah University of Science and Technology (KAUST). His current research interests include but not limited to 5G and beyond, aeronautical communications, wireless data centers, and underwater optical wireless communications, networking, and localization.
\end{IEEEbiography}

\begin{IEEEbiography}[{\includegraphics[width=1.1in,height=1.25in]{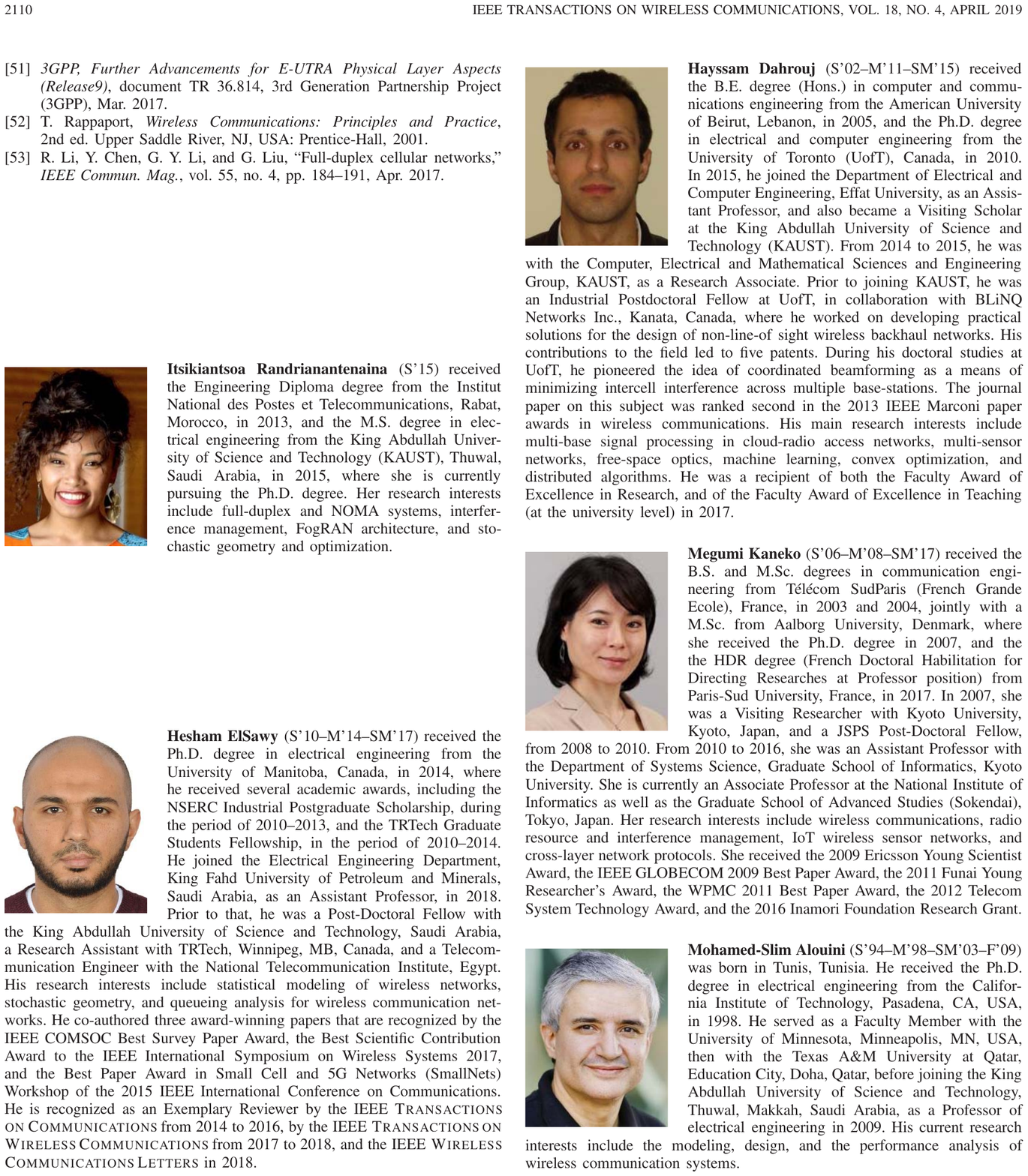}}]{Hesham ElSawy}
(S’10-M’14-SM’17) received the
Ph.D. degree in electrical engineering from the
University of Manitoba, Canada, in 2014, where
he received several academic awards, including the
NSERC Industrial Postgraduate Scholarship, during
the period of 2010–2013, and the TRTech Graduate
Students Fellowship, in the period of 2010–2014.
He joined the Electrical Engineering Department,
King Fahd University of Petroleum and Minerals,
Saudi Arabia, as an Assistant Professor, in 2018.
Prior to that, he was a Post-Doctoral Fellow with
the King Abdullah University of Science and Technology, Saudi Arabia, a Research Assistant with TRTech, Winnipeg, MB, Canada, and a Telecommunication Engineer with the National Telecommunication Institute, Egypt.
His research interests include statistical modeling of wireless networks, stochastic geometry, and queueing analysis for wireless communication networks. He co-authored three award-winning papers that are recognized by the IEEE COMSOC Best Survey Paper Award, the Best Scientific Contribution Award to the IEEE International Symposium on Wireless Systems 2017, and the Best Paper Award in Small Cell and 5G Networks (SmallNets)
Workshop of the 2015 IEEE International Conference on Communications. He is recognized as an Exemplary Reviewer by the \textsc{IEEE Transactions
On Communications} from 2014 to 2016, by the \textsc{IEEE Transactions On Wireless Communications} from 2017 to 2018, and the \textsc{IEEE Wireless Communications Letters} in 2018.\end{IEEEbiography}

\begin{IEEEbiography}[{\includegraphics[width=1.1in]{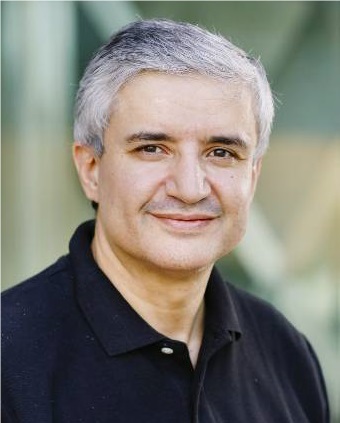}}]  {Mohamed-Slim Alouini} 	
	(S’94-M’98-SM’03-F’09)  was	
	born in Tunis, Tunisia. He received the Ph.D. degree in Electrical Engineering
	from the California Institute of Technology (Caltech), Pasadena,
	CA, USA, in 1998. He served as a faculty member in the University of Minnesota,
	Minneapolis, MN, USA, then in the Texas A\&M University at Qatar,
	Education City, Doha, Qatar before joining King Abdullah University of
	Science and Technology (KAUST), Thuwal, Makkah Province, Saudi
	Arabia as a Professor of Electrical Engineering in 2009. His current
	research interests include the modeling, design, and
	performance analysis of wireless communication systems.
\end{IEEEbiography}

\begin{IEEEbiography}[{\includegraphics[width=1in,height=1.25in]{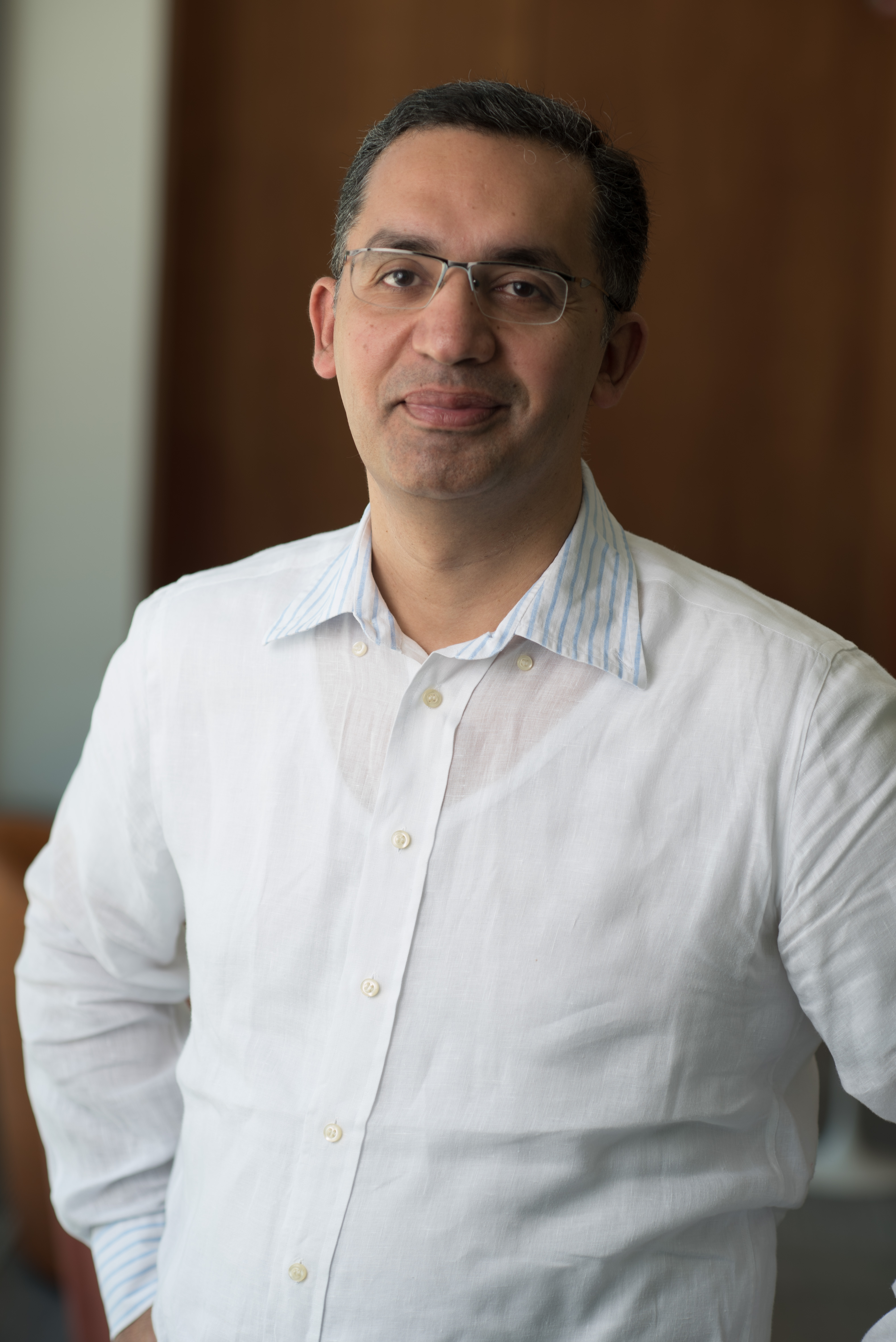}}]{Tareq Y. Al-Naffouri }
(M'10-SM'18) Tareq  Al-Naffouri  received  the  B.S.  degrees  in  mathematics  and  electrical  engineering  (with  first  honors)  from  King  Fahd  University  of  Petroleum  and  Minerals,  Dhahran,  Saudi  Arabia,  the  M.S.  degree  in  electrical  engineering  from  the  Georgia  Institute  of  Technology,  Atlanta,  in  1998,  and  the  Ph.D.  degree  in  electrical  engineering  from  Stanford  University,  Stanford,  CA,  in  2004.  

He  was  a  visiting  scholar  at  California  Institute  of  Technology,  Pasadena,  CA  in  2005  and  summer  2006.  He  was  a  Fulbright scholar  at  the  University  of  Southern  California  in  2008.  He  has  held  internship  positions  at  NEC  Research  Labs,  Tokyo,  Japan,  in  1998,  Adaptive  Systems  Lab,  University  of  California  at  Los  Angeles  in  1999,  National  Semiconductor,  Santa  Clara,  CA,  in  2001  and  2002,  and  Beceem  Communications  Santa  Clara,  CA,  in  2004.  He  is  currently  an  Associate Professor  at  the  Electrical  Engineering  Department,  King  Abdullah  University  of  Science  and  Technology  (KAUST).  His  research  interests  lie  in  the  areas  of  sparse, adaptive,  and  statistical  signal  processing  and  their  applications,  localization,  machine  learning,  and  network  information  theory.    He  has  over  240  publications  in  journal  and  conference  proceedings,  9  standard  contributions,  14  issued  patents,  and  8  pending. 

Dr.  Al-Naffouri  is  the  recipient  of  the  IEEE  Education  Society  Chapter  Achievement  Award  in  2008  and  Al-Marai  Award  for  innovative  research  in  communication  in  2009.  Dr.  Al-Naffouri  has  also  been  serving  as  an  Associate  Editor  of  Transactions  on  Signal  Processing  since  August  2013. 
\end{IEEEbiography}

\end{document}